\documentclass[11pt,a4paper]{article}

\usepackage{a4wide}

\usepackage{amsmath}
\usepackage{amssymb}
\usepackage{amsthm}
 
\usepackage{filecontents}

\newtheorem{theorem}{Theorem}
\newtheorem{lemma}[theorem]{Lemma}

\newtheorem{remark}[theorem]{Remark}
\newtheorem{definition}[theorem]{Definition}

\newtheorem*{thmmain}{Theorem \ref{thm:main}}
\newtheorem*{thmasymmain}{Theorem \ref{thm:asym-main}}
\newtheorem*{thmperfect}{Theorem \ref{thm:perfect}}
 
\usepackage{hyperref} 

\hypersetup{colorlinks=true,linkcolor=blue, citecolor=blue}

\newcommand{\ab}{\mathbf{a}}
\newcommand{\bb}{\mathbf{b}}
\newcommand{\cb}{\mathbf{c}}
\newcommand{\tb}{\mathbf{t}}
\newcommand{\xb}{\mathbf{x}}
\newcommand{\yb}{\mathbf{y}}
\newcommand{\zb}{\mathbf{z}}
\newcommand{\wb}{\mathbf{w}}

\newcommand{\ra}{\rightarrow}

\newcommand{\abs}[1]{\left\vert#1\right\vert}

\makeatletter
\def\prob#1#2#3{\goodbreak\begin{list}{}{\labelwidth\z@ \itemindent-\leftmargin
            \itemsep\z@  \topsep6\p@\@plus6\p@
            \let\makelabel\descriptionlabel}
        \item[\it Name]#1
        \item[\it Instance]                #2
        \item[\it Output]#3
    \end{list}}
    \makeatother

      \def\Rplus{{\mathbb R}_{\geq 0}}
     \def\condV{{\mathrm{cond}(\mathcal{V})}}
      
\def\calF{\mathcal{F}}
\def\calC{\mathcal{C}}
    
    \newcommand{\nSAT}{\#\mathsf{SAT}}
   \def\RHPi{\mathsf{\#RH}\Pi_1} 
    \def\numP{\#\mathsf{P}}    
    \def\RP{\mathsf{RP}}    
\def\FP{\mathsf{FP}}

    \def\HyperSpinf{\#{\mathsf{Multi2Spin}_{\Delta,c}}(f)  } 
    \def\HyperSpinfAP{\#{\mathsf{Multi2Spin}_{\Delta}}(f) } 
     
    \newcommand{\nCSP}[1]{\#{\mathsf{CSP}}(#1)}
        \newcommand{\nNRCSP}[1]{\#{\mathsf{NoRepeatCSP}}(#1)}
        \newcommand{\CSP}[1]{{\mathsf{CSP}}(#1)}
        \newcommand{\Holant}[1]{{\mathsf{Holant}}(#1)}
	\newcommand{\nCSPd}[1]{\#{\mathsf{CSP}_\Delta}(#1)}
		\newcommand{\nNRCSPd}[1]{\#{\mathsf{NoRepeatCSP}_\Delta}(#1)}
		\newcommand{\CSPd}[1]{{\mathsf{CSP}_\Delta}(#1)}
		\newcommand{\nCSPmyd}[2]{\#{\mathsf{CSP}_{#2}}(#1)}
	\newcommand{\nCSPdc}[1]{\#{\mathsf{CSP}_{\Delta,c}}(#1)} 
		\newcommand{\nNRCSPdc}[1]{\#{\mathsf{NoRepeatCSP}_{\Delta,c}}(#1)} 
    
    \def\eq{\mathsf{eq}}
    \def\EQ{\mathsf{EQ}}  
		\def\pinzero{\mathsf{pin}0}
		\def\pinone{\mathsf{pin}1}
    \def\NP{\mathsf{NP}} 
		\def\IM{IM_2} 
    \def\EASYk{\mathsf{EASY}(k)}     
    \def\NAND{\mathsf{NAND}}
    \def\Implies{\mathsf{Implies}}
    \def\Bor{\mathsf{OR}}
		\def\XOR{\mathsf{XOR}}

    \def\nBIS{\#\mathsf{BIS}} 
    \def\zeroes{\mathbf{0}}
    \def\ones{\mathbf{1}}
    
    \def\zerof{f_\mathsf{zero}}  
    \def\onef{f_\mathsf{one}}  
    \def\allzerof{f_\mathsf{allzero}}  
    \def\allonef{f_\mathsf{allone}}  
    \def\eqf{f_\mathsf{EQ}}  
    \def\evenf{f_\mathsf{even}}  
    \def\oddf{f_\mathsf{odd}}

\begin{document}
\title{Approximating partition functions of bounded-degree Boolean counting Constraint Satisfaction Problems\thanks
{To appear in JCSS. A preliminary announcement of these results  appeared in the proceedings of ICALP 2017.
The research leading to these results has received funding from the European Research Council under the European Union's Seventh Framework Programme (FP7/2007-2013) ERC grant agreement no.\ 334828. The paper reflects only the authors' views and not the views of the ERC or the European Commission. The European Union is not liable for any use that may be made of the information contained therein. Department of Computer Science, University of Oxford, Wolfson Building, Parks Road, Oxford, OX1~3QD, UK.}
}
\author{Andreas Galanis, Leslie Ann Goldberg, and Kuan Yang}

\date{20 August 2020} 
\maketitle

\begin{abstract}   
We study the complexity of $\nCSPd{\Gamma}$, which is the problem of counting satisfying assignments to CSP instances with constraints from $\Gamma$ and whose variables can appear at most $\Delta$ times. Our main result shows that: (i) if every function in $\Gamma$ is affine, then $\nCSPd{\Gamma}$ is in FP for all $\Delta$, (ii) otherwise, if every function in $\Gamma$ is in a class called $\IM$, then for large $\Delta$, $\nCSPd{\Gamma}$ is equivalent under approximation-preserving  reductions to the problem of counting independent sets in bipartite graphs, (iii) otherwise,  for large $\Delta$, it is $\NP$-hard to approximate $\nCSPd{\Gamma}$, even within an exponential factor.	
\end{abstract} 
    
\section{Introduction}
  
\emph{Constraint Satisfaction Problems} (CSPs), which originated in Artificial Intelligence~\cite{handbook}
provide a general framework for modelling decision, counting and approximate counting problems.
The paradigm is sufficiently general that applications from diverse areas such as
database theory, scheduling and graph theory
can all be captured  (see, for example, \cite{KV, Kumar, Montanari}).
Moreover, all graph homomorphism decision and counting problems~\cite{HN} can be re-cast in the CSP framework
and partition function problems from statistical physics~\cite{Welsh} 
can be represented as counting CSPs.
Given the usefulness of CSPs, the study of the complexity of CSPs is a an extremely active area in 
computational complexity (for example, see~\cite{dagstuhlreport} and the references therein).

In this paper, we will be concerned with  
Boolean counting CSPs. 
An instance $I=(V,\calC)$ of a Boolean counting CSP
consists of a set $V$ of \emph{variables} and a set $\calC$ of constraints.
An assignment $\sigma: V \rightarrow \{0,1\}$ assigns
a Boolean value called a ``spin'' to each variable.
Each constraint associates a tuple $(v_1,\ldots,v_k)$ of variables
with a Boolean relation which constrains the spins that
can be assigned to $v_1,\ldots,v_k$.
In particular, the  assignment $\sigma$  is said to ``satisfy'' the constraint
if the tuple $(\sigma(v_1),\ldots,\sigma(v_k))$ is in the corresponding relation.
An assignment is said to be ``satisfying'' if it satisfies all constraints.
A Constraint Satisfaction Problem comes with two important parameters --- 
the  constraint language $\Gamma$
is the set of all relations that may be used in constraints 
and the degree
$\Delta$ is the maximum number of times that any variable $v\in V$
may be used in constraints in any instance.
The number of satisfying assignments is denoted $Z_I$.
The computational problem $\nCSPd{\Gamma}$ is the problem of computing $Z_I$, given a CSP instance~$I$
with constraints in~$\Gamma$ and degree at most~$\Delta$.
We use $\nCSP{\Gamma}$ to denote the version of the problem in which the degree of instances is unconstrained.

Although constraints are supported by Boolean relations, they
can be used to code up weighted interactions such as those that arise in statistical physics.
For example, let $R$ be the ``not-all-equal'' relation of arity~$3$.
Then consider the conjunction of $R(x,a,b)$ and $R(y,a,b)$.
There  are two satisfying assignments with $\sigma(x)=0$ and $\sigma(y)=1$ since $\sigma(a)$ and $\sigma(b)$ 
must differ. Similarly, there are two satisfying assignments with $\sigma(x)=1$ and $\sigma(y)=0$.
On the other hand, there are three satisfying assignments with $\sigma(x)=\sigma(y)=1$
and there are three satisfying assignments with $\sigma(x)=\sigma(y)=0$.
Thus, the induced interaction on the variables $x$ and $y$ 
is the same as the interaction of the ferromagnetic Ising model (at an appropriate temperature) --- 
an assignment in which $x$ and $y$ have the same spin has weight~$3$, whereas an assignment
where they have different spins has weight~$2$.

For every $\Delta \geq 3$, the work of Cai, Lu and Xia \cite{CLX} completely classifies 
the complexity of exactly solving $\nCSPd{\Gamma}$, depending on the parameter~$\Gamma$.
If every relation in $\Gamma$ is affine, then $\nCSPd{\Gamma}$ is solvable in polynomial time
(so the problem in the complexity class $\FP$).
Otherwise, it is $\numP$-complete.
The term ``affine'' will be defined in Section~\ref{sec:prelim}. Roughly, it means that the tuples in the relation
are solutions to a linear system, so Gaussian elimination gives an appropriate polynomial-time algorithm.
The characterisation of Cai, Lu and Xia
 is exactly the same classification that was obtained for the unbounded problem $\nCSP{\Gamma}$
by Creignou and Hermann~\cite{CH}. Thus, as far as exact counting is concerned, the
degree-bound~$\Delta$ does not affect the complexity as long as $\Delta \geq 3$. 
As Cai, Lu and Xia point out, the dichotomy is false
for $\Delta=2$, where $\nCSPmyd{\Gamma}{2}$ is equivalent to the Holant problem $\Holant{\Gamma}$ ---
see the references in~\cite{CLX} for more information about Holant problems.

Much less is known about the complexity of \emph{approximately} solving $\nCSPd{\Gamma}$.
In fact, even the \emph{decision problem} is still open.
While Schaefer~\cite{Schaefer} completely classified the complexity of the decision problem 
$\CSP{\Gamma}$ --- where the goal is to determine whether or not $Z_I$ is~$0$ for an instance of~$\nCSP{\Gamma}$ ---
the complexity of the corresponding decision problem $\CSPd{\Gamma}$, where the instance has degree at most~$\Delta$,
is still not completely resolved. For $\Delta\geq 3$, Dalmau and Ford~\cite{DF} have solved the special
case where $\Gamma$ includes both of the relations
$R_{\delta_0} = \{0\}$ and $R_{\delta_1} = \{1\}$.
This special case is known as the ``conservative case'' in the CSP literature.
For $\Delta \geq 6$, Dyer et al.~\cite{DGJR} 
have classified the difficulty of the approximation problem:
\begin{itemize}
\item If every relation in $\Gamma$ is affine, then $\nCSPd{\Gamma \cup \{R_{\delta_0},R_{\delta_1}\}}$ is in~$\FP$.
\item Otherwise, if every relation in $\Gamma$ is in a class called $\IM$ (a class which will be defined in Section~\ref{sec:prelim})
then $\nCSPd{\Gamma \cup \{R_{\delta_0},R_{\delta_1}\}}$
is equivalent 
under approximation-preserving (AP) reductions to   the
counting problem~$\nBIS$ (the problem of counting independent sets in bipartite graphs).
\item Otherwise, there is no FPRAS for 
$\nCSPd{\Gamma \cup \{R_{\delta_0},R_{\delta_1}\}}$
unless $\NP=\RP$.
\end{itemize}
Dyer et al.\ made only partial progress on the cases where $\Delta \in \{3,4,5\}$.
We refer the reader to~\cite{DGJR, LiuLuMonotone} for
a discussion of the partial classification. However, it is worth noting here
that the complexity of $\nCSPd{\Gamma \cup \{R_{\delta_0},R_{\delta_1}\}}$
is closely related to the complexity of counting satisfying assignments of so-called
read-$d$ Monotone CNF Formulas.  Crucial progress was made by Liu and Lu \cite{LiuLuMonotone},
who completely resolved the complexity of the latter problem.
Given the work of Liu and Lu, a complete classification of $\nCSPd{\Gamma \cup \{R_{\delta_0},R_{\delta_1}\}}$
for $\Delta \in \{3,4,5\}$ may be in reach.

The restriction that $R_{\delta_0}$ and $R_{\delta_1}$ are contained in $\Gamma$
is a severe one because it does not apply to many natural applications.
On the other hand, we are a long way from a precise understanding of the 
complexity of $\nCSPd{\Gamma}$ without this restriction because there are specific, relevant
parameters that we do not understand.
For example, for a positive integer~$k$, let $\Gamma$ 
be the singleton set containing only
 the arity-$k$ ``not-all-spin-1''
 relation. Then satisfying assignments of an instance of $\nCSPd{\Gamma}$ correspond to
independent sets of a $k$-uniform hypergraph with maximum degree $\Delta$. The current state-of-the-art for this problem is that there
is an FPRAS for $\Delta=O(2^{k/2})$ \cite{HSZ} and that the problem is NP-hard to approximate for $\Delta=\Omega(2^{k/2})$ \cite{HyperIS}; the implicit constants in these bounds do not currently match and thus, for large $k$, there is a large range of $\Delta$'s where we do not yet know the complexity of approximating $\nCSPd{\Gamma}$.
If $\Gamma$ instead contains (only) 
the arity-$k$ ``at-least-one-spin-0''  relation
then satisfying assignments of an instance of 
$\nCSPd{\Gamma}$ correspond to the so-called
``strong'' independent sets of a $k$-uniform hypergraph.
Song, Yin and Zhao~\cite{SYZ} 
have presented a barrier for hardness results, showing
why  current technology
is unsuitable for resolving the cases where $\Delta \in \{4,5\}$ (roughly, these cases are in 
 ``non-uniqueness'', but this is not realisable by finite gadgets).

The purpose of the present paper is to remove the severe restriction 
that $R_{\delta_0}$ and $R_{\delta_1}$ are contained in $\Gamma$
in the approximate counting classification of $\nCSPd{\Gamma}$
from~\cite{DGJR}.
Since pinning down precise thresholds seems a long way out of reach, 
we instead focus on whether there is a ``barrier'' value $\Delta_0$ such that, for all $\Delta \geq \Delta_0$,
approximation is intractable.
Since we wish to get the strongest possible inapproximability results
(showing the hardness of approximating $Z_I$ even within an exponential factor), we 
define the following computational problem, which has an extra parameter $c>1$ that
captures the desired accuracy.
\prob{ $\nCSPdc{\Gamma}$.}
{ An $n$-variable instance $I$ of a CSP with constraint language~$\Gamma$ and degree at most~$\Delta$.}
{  A number $\widehat{Z}$ such that $c^{-n}Z_{I}\leq \widehat{Z}\leq c^nZ_{I}$.}

Although we have not yet defined
all of the terms, we can now at least state (a weak version of) our result.

\newcommand{\statethmmainrelversion}{Let $\Gamma$ be a 
Boolean constraint language. Then,
\begin{enumerate}
\item If every  relation in $\Gamma$ is affine then $\nCSP{\Gamma}$ is in $\FP$. 
\item Otherwise, if every relation in $\Gamma$ is in the class $\IM$,  then there exists an integer~$\Delta_0$ such that for all $\Delta\geq \Delta_0$, $\nCSPd{\Gamma}$ is $\nBIS$-equivalent under $\mathsf{AP}$-reductions.
\item Otherwise, there exists an integer~$\Delta_0$ such that for all $\Delta\geq \Delta_0$, there exists 
a real number~$c>1$ such that $\nCSPdc{\Gamma}$ is $\NP$-hard.
\end{enumerate}}   

\newcommand{\statethmmain}{Let $\Gamma$ be a 
Boolean constraint language. Then,
\begin{enumerate}
\item If every function in $\Gamma$ is affine then $\nCSP{\Gamma}$ 
and $\nNRCSP{\Gamma}$ are 
both in $\FP$. 
\item Otherwise, if $\Gamma\subseteq IM_2$, then there exists an integer~$\Delta_0$ such that for all $\Delta\geq \Delta_0$, 
 $\nCSPd{\Gamma}$ 
and $\nNRCSPd{\Gamma}$ are both
 $\nBIS$-equivalent under $\mathsf{AP}$-reductions,  and
\item Otherwise, there exists an integer~$\Delta_0$ such that for all $\Delta\geq \Delta_0$, there exists 
a real number~$c>1$ such that 
$\nCSPdc{\Gamma}$ and 
$\nNRCSPdc{\Gamma}$ are both $\NP$-hard.
\end{enumerate}}  
\begin{theorem}\label{thm:main-simplified}
\statethmmainrelversion{}
\end{theorem}

After defining all of the terms, we will  state a stronger theorem, Theorem~\ref{thm:main},
which immediately implies Theorem~\ref{thm:main-simplified}.
The stronger version applies to the \#CSP problems that we have already introduced,
but it also applies to other restrictions of these problems,
which have even more applications.

We now explain the restriction.
Note that in the CSP framework, as we have defined it, the variables that are constrained by a 
given constraint need not be distinct.
Thus, if the arity-$4$  relation~$R$ is present in a constraint language~$\Gamma$,
then an instance  of $\nCSP{\Gamma}$ 
with variables~$x$ and~$y$
may contain
a constraint such as $R(x,x,y,x)$.
This ability to repeat variables is equivalent to assuming that 
equality relations of all arities are present in~$\Gamma$.
This feature of the CSP definition is inconvenient for two reasons:
(1) It does not fit well with some spin-system applications, and
(2) In many settings, it obscures 
the   nuanced complexity classification that  arise.

As an example of (1), recall   the  application 
where  $\Gamma$ is the singleton set
containing only  the arity-$k$   ``not-all-spin-1'' relation.  As we noted earlier,
satisfying assignments of a $\nCSP{\Gamma}$ instance correspond to
independent sets of a $k$-uniform hypergraph.
Here, hyperedges are size-$k$ subsets of vertices and it does not
make sense to allow repeated vertices!

The point (2) is well-known. In fact, the ``equality is always present'' assumption is
the main feature that separates \#CSPs from the more general Holant framework~\cite{Caisurvey}.
 
In our current setting, 
it turns out that adding equality functions to~$\Gamma$ does not change the complexity classification,
but this is a result of our theorems rather than an a priori assumption --- indeed,
determining  which constraint languages $\Gamma$ can appropriately simulate
equality functions is  one of the difficulties --- thus, throwing equalities in ``for free''
would  substantially weaken our results!
Our main result, Theorem~\ref{thm:main}, which will be presented in Section~\ref{sec:results},
applies both to the \#CSPs that we have already defined, and to
more refined versions, in which constraints may not repeat variables.

We wish now to discuss an important special case
in which both the \#CSPs and the refined versions have already been studied.
This is the special case in which $\Gamma$ consists of a single relation
which is symmetric in its arguments. 
A symmetric relation that is not affine is not in $\IM$. Therefore, 
Item~2 in the statement of  Theorem~\ref{thm:main-simplified}
never arises in this special case. Our earlier paper~\cite{symmetric} 
shows that, in this case (where $\Gamma$ consists of a single, symmetric, non-affine relation) 
 there  is an integer~$\Delta_0$ such that for all $\Delta\geq \Delta_0$, there exists 
a real number~$c>1$ such that $\nCSPdc{\Gamma}$ is $\NP$-hard.

While the work of~\cite{symmetric} is important for this paper, note that
the special case is far from general --- in particular, it is easy to  induce asymmetric constraints using symmetric ones.
For example, suppose that $R_1$ is the (symmetric)
arity-2 ``not-all-spin-1'' constraint, $R_2$ is the (symmetric) arity-2   ``not the same spin''
constraint and $R_3=\{(0,0),(0,1),(1,1)\}$ is the (asymmetric) arity-2   ``Implies'' constraint.
Then the conjunction of $R_1(x,a)$ and $R_2(a,y)$ induces $R_3(x,y)$.

It is interesting that Theorem~\ref{thm:main-simplified} 
is exactly the same classification that was obtained for the \emph{unbounded} problem $\nCSP{\Gamma}$ by
Dyer et al.~\cite{trichotomy}. 
In particular, they showed  
\begin{enumerate}
 \item If every  relation in $\Gamma$ is affine then $\nCSP{\Gamma}$ is in $\FP$. 
\item Otherwise, if every relation in $\Gamma$ is in the class $\IM$,  then 
$\nCSP{\Gamma}$ is 
  $\nBIS$-equivalent under $\mathsf{AP}$-reductions.
\item Otherwise, $\nCSP{\Gamma}$ is $\nSAT$-equivalent under 
$\mathsf{AP}$-reductions, where $\nSAT$ is the problem of  counting the satisfying assignments of 
a Boolean formula. 
\end{enumerate}
The   inapproximability that we demonstrate in Item~3 of Theorem~\ref{thm:main-simplified} is stronger
than what was known in the unbounded case, both (obviously) because of the degree bound, but
also because we show that it is hard to get within an exponential factor.
(This strong kind of inapproximability was also missing from the results of \cite{DGJR}).

\section {Definitions and Statement of Main Result}\label{sec:properties}\label{sec:prelim} \label{sec:results}

Before giving formal definitions of the problems that we study, we introduce some notation.
We use boldface letters to denote Boolean vectors. 
A \emph{pseudo-Boolean} function is a function of the form
$f: \{0, 1\}^k \rightarrow \Rplus$ 
for some positive integer~$k$, which is called the \emph{arity} of~$f$.
\begin{definition} Given a pseudo-Boolean function $f: \{0, 1\}^k \rightarrow \Rplus$ ,
we use the notation $R_f$ to denote the relation 
$R_f=\{\xb\in\{0,1\}^k\mid f(\xb)>0\}$, which is the relation underlying~$f$.
\end{definition}
If the range of $f$ is $\{0,1\}$ then $f$ is said to be a \emph{Boolean function}
and of course in that case $R_f=\{\xb\in\{0,1\}^k\mid f(\xb)=1\}$.

In order to allow consistency with obvious generalisations, our formal definition of
the Boolean Constraint Satisfaction Problem is in terms of  
 Boolean functions  
 (rather than,
equivalently,  using the underlying relations).

A \emph{Constraint language} $\Gamma$ is a set of
pseudo-Boolean functions. It is a \emph{Boolean constraint language} if all of the functions in it
are Boolean functions. 
An instance~$I=(V,\calC)$ of 
a CSP with constraint language~$\Gamma$
 consists of a set $V$ of variables and a set $\calC$ of constraints.
Each constraint $C_i \in \calC$ is of the form $f_i(v_{i,1},\ldots,v_{i,k_i})$ 
where $f_i$ is an arity-$k_i$   function in~$\Gamma$ 
and $(v_{i,1},\ldots,v_{i,k_i})$ is a tuple of (not necessarily distinct)  variables in~$V$.
The constraint $C_i$ is said to be ``Repeat-Free'' if all of the variables are distinct.
Each 
\emph{assignment} $\sigma: V\rightarrow \{0,1\}$ of Boolean values to the variables in~$V$
has a weight
$$w_{I}(\sigma):=\prod_{f_i(v_{i,1},\hdots,v_{i,k_i})\in\calC} f_i(\sigma(v_{i,1}),\hdots,\sigma(v_{i,k_i})).$$
The \emph{partition function} maps the instance~$I$ 
to the quantity 
$$Z_{I}:=\sum_{\sigma:V\rightarrow\{0,1\}}w_{I}(\sigma)=\sum_{\sigma:V\rightarrow\{0,1\}}
\prod_{f_i(v_{i,1},\hdots,v_{i,k_i})\in \calC} f_i(\sigma(v_{i,1}),\hdots,\sigma(v_{i,k_i})).$$ 

If $\Gamma$ is a Boolean constraint language
then  it is easy to see that $w_I(\sigma) = 1$
if the assignment is satisfying and $w_I(\sigma)=0$, otherwise. 
Thus,  $Z_I$ is the number of
satisfying assignments of~$I$.

When $Z_{I}>0$, we will use $\mu_{I}(\cdot)$ to denote the Gibbs distribution corresponding to~$Z_{I}$. This is the probability distribution on the set of assignments $\sigma:V\rightarrow\{0,1\}$ such that  
\[\mu_{I}(\sigma)=\frac{w_{I}(\sigma)}{Z_{I}}\mbox{ for all } \sigma:V\rightarrow\{0,1\}.\]

The \emph{degree}  $d_v(C)$ of a variable~$v$ in a constraint~$C$
is the number of times that the variable~$v$ appears in the tuple corresponding to~$C$ and the
degree $d_v$ of the variable is $d_v = \sum_{C \in \calC} d_v(C)$.
Finally, the degree of the instance $I$ is $\max_{v\in V} d_v$.

\begin{definition}
$\nCSPd{\Gamma}$ is the problem of computing $Z_I$, given a CSP instance~$I$
with constraints in~$\Gamma$ and degree at most~$\Delta$.
$\nCSP{\Gamma}$
is the   version of the problem in which the degree of instances is unconstrained.
$\nCSPdc{\Gamma}$  has an extra parameter $c>1$ that 
captures the desired accuracy.
The problem is to compute a number $\widehat{Z}$ such that 
$c^{-n} Z_I \leq \widehat{Z} \leq c^n Z_I$, where $n$ is the number of variables in the instance~$I$.
The problems $\nNRCSPd{\Gamma}$, $\nNRCSP{\Gamma}$ and
$\nNRCSPdc{\Gamma}$ are defined similarly, except that inputs are restricted so that
all constraints are Repeat-Free.
\end{definition}

\begin{definition}\label{def:affine}
A Boolean function $f:\{0,1\}^k\ra \{0,1\}$ is \emph{affine}
if there is a $k\times k$ Boolean matrix~$\mathbf{A}$ and a length-$k$ Boolean vector $\mathbf{b}$
such that $R_f$ is equal to the set of solutions~$\xb$ of  $\mathbf{A}\xb=\mathbf{b}$ over $\mathrm{GF}(2)$.
\end{definition}

\begin{definition}[The set of functions $IM_2$]\label{def:IM2}
A Boolean function $f:\{0,1\}^k\rightarrow\{0,1\}$ is in $\IM$ if $f(x_1,\hdots,x_k)$ is logically equivalent to a conjuction of (any number of) predicates of the form $x_i$, $\neg x_i$ or $x_i\Rightarrow x_j$.
\end{definition}

We have now defined all of the terms  in our main theorem
apart from  some well-known concepts from complexity theory, which we discuss next.
$\FP$ is the class of computational problems (with numerical output) that can be solved in polynomial time.
An FPRAS is a randomised algorithm that 
produces approximate solutions within specified relative error with high probability in 
polynomial time.  
For two counting problems $\#\mathsf{A}$ and $\#\mathsf{B}$, we say that $\#\mathsf{A}$ is $\#\mathsf{B}$-easy 
if there is an approximation-preserving (AP)-reduction
from~$\#\mathsf{A}$ to $\#\mathsf{B}$. The formal definition of an AP-reduction
can be found in~\cite{DGGJ}. It is
a randomised Turing reduction that yields close approximations to 
$\#\mathsf{A}$ when provided with close approximations to~$\#\mathsf{B}$.
The definition of AP-reduction meshes with the definition of FPRAS in the sense
that the existence of an FPRAS  for $\#\mathsf{B}$ implies the existence of an FPRAS for $\#\mathsf{A}$. 
We say that $\#\mathsf{A}$ is $\#\mathsf{B}$-hard if 
there is an AP-reduction from~$\#\mathsf{B}$ to $\#\mathsf{A}$. 
  Finally, we say that $\#\mathsf{A}$ is $\#\mathsf{B}$-equivalent if $\#\mathsf{A}$ is both $\#\mathsf{B}$-easy and $\#\mathsf{B}$-hard.

The problem of counting satisfying assignments 
of a Boolean formula is denoted by $\nSAT$. 
Every counting problem in $\numP$ is AP-reducible to $\nSAT$, so $\nSAT$ is said to be
complete for $\numP$ with respect to AP-reductions. It is known that there
is no FPRAS for $\nSAT$ unless $\RP=\NP$.
The problem of counting independent 
sets in a bipartite graph is denoted by $\nBIS$.  The problem $\nBIS$ appears to be
of intermediate complexity:  there is no known FPRAS for $\nBIS$ (and it is generally believed
that none exists) but there is no known AP-reduction from 
$\nSAT$ to $\nBIS$. Indeed, $\nBIS$ is complete with respect to AP-reductions for
a complexity class $\RHPi$.

Given all of these definitions, we now formally state the stronger version of Theorem~\ref{thm:main-simplified} promised in the introduction. The proof can be found in Section~\ref{sec:proofofmain}.
\begin{theorem}\label{thm:main}
\statethmmain{}
\end{theorem} 

\section{Overview of the Proof of Theorem~\ref{thm:main}}\label{sec:sketcha}
 
In this section, we give a non-technical overview of 
the proof of Theorem~\ref{thm:main}.
Our objective is to illustrate the main ideas and obstacles without delving into the  more detailed definitions. A more technical overview can be found in Section~\ref{sec:sketchb}. Our focus in this section will be on the case where $\Gamma$ consists of a single Boolean function $f:\{0,1\}^k\rightarrow \{0,1\}$. 
As will be clear in Section~\ref{sec:proofofmain}, this case is the main ingredient in the proof of the theorem.

A typical approach for showing that a counting CSP 
is intractable is to 
use an instance of the CSP
to build a ``gadget'' 
which  simulates 
an intractable  binary 2-spin constraint.
This was the approach used in~\cite{symmetric}, which  proved the  intractability of 
$\nNRCSPd{\{f\}}$ for any \emph{symmetric} non-affine Boolean function~$f$
by constructing an instance~$I$
of $\nNRCSPd{\{f\}}$, 
along with variables~$x$ and~$y$,
such that 
for all spins $s_x\in \{0,1\}$ and $s_y\in\{0,1\}$
the marginal distribution
$\mu_I(x,y)$  
satisfies
\begin{equation}
\label{eq:examplesim}
\mu_I(\sigma(x)=s_x, \sigma(y) = s_y) = \frac{g(s_x,s_y)}{g(0,0)+g(0,1)+g(1,0)+g(1,1)},
\end{equation}
where $g$ is a binary function  that codes up the interaction   of
an  intractable anti-ferromagnetic 2-spin system.
We will not need to give  detailed definitions of 2-spin systems in this paper. Instead, we
give a sufficient condition for intractability.

\begin{definition}\label{def:hard-function}
A binary function $g:\{0, 1\}^2 \ra \Rplus$ is said to be ``\emph{hard}'' if all of the following hold:
\begin{align*}
&\, g(0,0) + g(1, 1) > 0,\\
&\min\{g(0, 0), g(1, 1)\} < \sqrt{g(0, 1)g(1, 0)},\\ 
&\max\{g(0, 0), g(1, 1)\} \leq \sqrt{g(0, 1)g(1, 0)}.
\end{align*}
\end{definition}

It was established in~\cite{symmetric} that the ability to ``simulate'' a hard function~$g$
in the sense of~\eqref{eq:examplesim}
ensures that 
$\nNRCSPd{\{f\}}$  
is $\NP$-hard to approximate, even within an exponential factor.

A key feature of \emph{symmetric} Boolean functions~$f$ which 
facilitated such simulation in~\cite{symmetric} 
 was the fact that the class of relevant hard functions~$g$
 is well-behaved, and  it turned out
 that  
 it suffices to encode such a hard binary function with only $\epsilon$-accuracy, for some sufficiently small $\epsilon> 0$, and 
 this was enough to ensure the $\NP$-hardness of
  $\nCSPdc{\{f\}}$.

The main obstacle in adapting the approach of \cite{symmetric} 
to the case where $f$ need not be symmetric in its arguments
arises when $f$ is in $\IM$. 
It is unlikely that such a function~$f$ can simulate a hard function $g$ in the sense of~\eqref{eq:examplesim} ---
indeed such a simulation would prove the (very surprising) result that  $\nBIS$ does not have an FPRAS (unless $\NP=\RP$). 
Thus, for $f\in \IM$, we need  instead to encode a binary function which will allow  
us to connect the problem $\nNRCSPd{\{f\}}$  
  to $\nBIS$.  

Now consider the binary Boolean function $\Implies$  
whose underlying relation  $R_{\Implies} = \{(0,0),(0,1),(1,1)\}$
contains all $(x,y)$ satisfying $x\Rightarrow y$.
Obviously, $\Implies$ is not symmetric, and it is not hard according 
to Definition~\ref{def:hard-function}.
On bipartite instances, however, the symmetry can be restored by interpreting differently the spins 0 and 1 on the two parts of the graph, and this leads to  a connection with $\nBIS$.
In particular, it is well-known~\cite{trichotomy} that $\nCSP{\{\Implies\}}$ is equivalent to $\nBIS$ under AP-reductions.   
This connection was extended to the bounded-degree setting by~\cite{BoundedBIS}.

Unfortunately, the symmetrisation which connects $\nCSP{\{\Implies\}}$ to $\nBIS$ is not very robust.
For example, suppose that a (non-symmetric) Boolean function~$f$
can be used to simulate, in the sense of~\eqref{eq:examplesim}, a binary function $g$
which is very close to~$\Implies$.
In particular, suppose that for some $\epsilon>0$ and
$\epsilon_1,\epsilon_2,\epsilon_3,\epsilon_4$ satisfying
$|\epsilon_i|\leq \epsilon$ for $i=1,2,3,4$, we have
\[\begin{array}{ll} g(0,0)=1+\epsilon_1, & g(0,1)=1+\epsilon_2,\\  g(1,0)=\epsilon_3, & g(1,1)=1+\epsilon_4.\end{array}\]
Such a close approximation is about the best that can be expected using 
the kind of approximate encodings that are available. 
However, 
the complexity of 
asymmetric 2-spin systems is not sufficiently well understood to exploit such a simulation.
Surprisingly, for \emph{any} arbitrarily small constant $\epsilon>0$,  it is not known even whether the unbounded degree version $\nCSP{\{g\}}$ is $\nBIS$-hard, 
and certainly nothing is known in our bounded-degree setting! 
 The trouble is that the symmetrisation that works for $\Implies$ (i.e., when $\epsilon_i=0$ for $i=1,2,3,4$) is no longer guaranteed to symmetrise the imperfect version with the $\epsilon_i$'s, 
 so the swapping of spin-0 and spin-1 values on one side of the bipartite 
 graph leads to an \emph{asymmetric}  2-spin system on bipartite graphs 
 and this does not fall  into the scope of known results 
  \cite{BoundedBIS} concerning bounded-degree bipartite 2-spin systems.

Our approach to handle this problem  for $f\in \IM$ is to carefully ensure that there is no accuracy error $\epsilon$ in encoding the function $\Implies$. In other words,  we show that, using $f\in \IM$, we can encode $\Implies$ \emph{perfectly}, a task which  is surprisingly intricate in the repeat-free setting. Our main technical theorem, 
 Theorem~\ref{thm:asym-main},  achieves  this goal.
Namely,  it shows that, for every non-affine Boolean function $f$, either $f$ simulates a hard function
(with arbitrarily small accuracy-error $\epsilon$,
which leads to the desired intractability of $\nNRCSPd{\{f\}}$)
or 
else $f$ ``supports perfect equality'' --- a concept which 
will be defined later, but essentially means that $f$ can be used
to perfectly simulate  the binary function $\EQ$ with underlying relation $R_{\EQ} = \{(0,0),(1,1)\}$.
Using $\EQ$, it is possible to simulate repeated variables in constraints,
so the  $\nBIS$-hardness of $\nCSPd{\{f\}}$ follows from \cite{trichotomy}.
When $f\notin \IM$
but $f$ supports perfect equality,
instead of reducing to the work in~\cite{trichotomy}, we 
 work somewhat harder to make sure that we also get the strong 
(exponential factor) 
inapproximability given in Theorem~\ref{thm:main}.

\section{Pinning, equality and simulating functions}\label{sec:pineqsim}

We will often be interested in the case where $\Gamma$ contains a single  function~$f:\{0,1\}^k\rightarrow \Rplus$.
In this case, we  can we simplify the notation because the constraints
in an instance~$I$ are in one-to-one correspondence with $k$-tuples of variables (there is no need to
repeat the name of the function~$f$ in each constraint).
So, for convenience, we make the following definitions.

A \emph{$k$-tuple hypergraph} $H=(V,\calF)$  
consists of 
a set $V$ of vertices, together with a set $\calF$ of \emph{hyperarcs}, where
every hyperarc in~$\calF$ is a $k$-tuple of distinct vertices in~$V$. 
The degree of $H$ is the maximum, over all vertices $v\in V$,
of the number of hyperarcs that contain~$v$.
Given a   function~$f: \{0,1\}^k \rightarrow \Rplus$,
we let $I_f(H)$ denote the instance of $\nNRCSP{\{f\}}$
whose constraints correspond to the hyperarcs of~$H$.
Given an assignment $\sigma\colon V \to \{0,1\}$  
we define
$w_{f;H}(\sigma):=\prod_{(v_1,\hdots,v_k)\in\mathcal{F}} f(\sigma(v_1),\hdots,\sigma(v_k))$
and  
$Z_{f;H}:=\sum_{\sigma:V\rightarrow\{0,1\}}w_{f;H}(\sigma),$ so
$Z_{I_f(H)} = Z_{f;I_f(H)}$.
By analogy to the Gibbs distribution on satisfying assignments,
when $Z_{f;H}>0$, we  use $\mu_{f;H}(\cdot)$ to denote the   probability distribution 
in which, for all assignments $\sigma:V\rightarrow\{0,1\}$,
$\mu_{f;H}(\sigma)={w_{f;H}(\sigma)}/{Z_{f;H}}$.
Given a  function $f:\{0,1\}^k \rightarrow \Rplus$,
a positive integer~$\Delta$, and a real number $c>1$, the
 following computational problems are equivalent to
 $\nNRCSPd{\{f\}}$ and $\nNRCSPdc{\{f\}}$, respectively.

\prob{ $\HyperSpinfAP$.}
{ A $k$-tuple  hypergraph $H$ with degree at most $\Delta$.}
{ The partition function $Z_{f;H}$.}

\prob{ $\HyperSpinf$.}
{ An $n$-vertex $k$-tuple  hypergraph $H$ with degree at most $\Delta$. }
{  A number $\widehat{Z}$ such that $c^{-n}Z_{f;H}\leq \widehat{Z}\leq c^nZ_{f;H}$.}

The name $\HyperSpinfAP$ 
indicates that the problem is to compute the partition function of
a 2-spin system with multi-body interactions 
specified by~$f$ and degree-bound~$\Delta$.

\subsection{Supporting pinning and equality}\label{sec:pinning}
Let $k$ be a positive integer and let
$H=(V,\mathcal{F})$ be a  $k$-tuple  hypergraph. Given a configuration $\sigma:V\rightarrow \{0,1\}$ and a subset $T\subseteq V$, we will use $\sigma_T$ to denote the restriction of~$\sigma$ to vertices in $T$. For a vertex $v\in V$, we will also use $\sigma_v$ to denote the spin $\sigma(v)$ of vertex $v$ in $\sigma$. The following definitions are 
generalisations of definitions from~\cite{symmetric}.

\begin{definition}\label{def:pinning}
Let $f:\{0,1\}^k\rightarrow\Rplus$. Suppose that $\epsilon\geq 0$ and $s\in \{0,1\}$. The 
$k$-tuple  hypergraph $H$ is an \emph{$\epsilon$-realisation of  pinning-to-$s$} if there exists a vertex $v$ of $H$ such that $\mu_{f;H}(\sigma_v=s)\geq 1-\epsilon$.  
\end{definition}

\begin{definition}\label{def:boundedsupport1}
Let $f:\{0,1\}^k\rightarrow\Rplus$ and
$s\in\{0,1\}$. We say that  $f$ supports \emph{pinning-to-$s$}   if, for every $\epsilon>0$, there is a  
 $k$-tuple hypergraph which is an  $\epsilon$-realisation of  pinning-to-$s$. 
We say that $f$
 supports \emph{perfect} pinning-to-$s$ if there 
 is a $k$-tuple hypergraph   which is a $0$-realisation of  pinning-to-$s$.
\end{definition}

We now define what it means for a function $f$ to support (perfect) equality which was already discussed in Section~\ref{sec:sketcha}.   
\begin{definition}\label{def:equality}
Let $f:\{0,1\}^k\rightarrow\Rplus$ and
$\epsilon\geq 0$. The  $k$-tuple hypergraph $H$ is an \emph{$\epsilon$-realisation of equality} if there 
exist distinct vertices $v_1$ and $v_2$ of~$H$ such that, for each $s\in\{0,1\}$,
\[\mu_{f;H}(\sigma_{v_1}=\sigma_{v_2}=s)\geq (1-\epsilon)/2.\]
\end{definition}
     
\begin{definition}\label{def:boundedsupport2a}
Let $f:\{0,1\}^k\rightarrow\Rplus$. The
function  $f$ \emph{supports equality} if, for every $\epsilon>0$, there is a  $k$-tuple hypergraph   which is an $\epsilon$-realisation of equality. The
function  $f$ \emph{supports perfect equality} if there is a $k$-tuple hypergraph which is a $0$-realisation of equality.
\end{definition}

\subsection{Realising conditional distributions induced by pinning and equality}\label{sec:conddistrib}

Given a set~$S$ of vertices,  it will be convenient to 
follow~\cite{symmetric} as follows. We
write $\sigma_S=\zeroes$ to denote the event that all vertices in $S$ are assigned the spin~$0$ under the assignment~$\sigma$.  We  similarly write $\sigma_S=\ones$
to denote the event that all vertices in $S$ are assigned the spin~$1$ under the assignment~$\sigma$.  
Finally, we use $\sigma^{\eq}_S$ to denote the event that all vertices in $S$ have the same spin under $\sigma$ (the spin could be~$0$ or~$1$). The following definition is a generalisation of Definition~16 of~\cite{symmetric}
except that we have changed the notation slightly for convenience.

\begin{definition}[{\cite[Definition 16]{symmetric}}] \label{def:generalpinfull}
Let $f:\{0,1\}^k\rightarrow\Rplus$. Let $H=(V,\mathcal{F})$ be a  $k$-tuple hypergraph. Let  $\mathcal{V}=(V_{\pinzero},V_{\pinone},\mathcal{V}_{\eq})$ where $V_{\pinzero}$ and $V_{\pinone}$ are disjoint subsets of $V$ and $\mathcal{V}_{\eq}$ is a (possibly empty) set of disjoint subsets of $V\backslash(V_{\pinzero}\cup V_{\pinone})$. Suppose that: (i) $V_{\pinzero}=\emptyset$ if $f$ does not support pinning-to-0, (ii) $V_{\pinone}=\emptyset$ if $f$ does not support pinning-to-1, (iii) $\mathcal{V}_{\eq}=\emptyset$ if $f$ does not support equality, (iv) it holds that 
$\mu_{f;H}(\sigma_{V_{\pinzero}}=\mathbf{0}, \sigma_{V_{\pinone}}=\mathbf{1}, \bigcap_{W\in \mathcal{V}_{\eq}}\sigma_{W}^{\eq})>0$. 
We will then say that ``$\mathcal{V}$ is \emph{admissible}  for  $H$ with respect to~$f$''   and we will denote by $\mu_{f;H}^{\condV}$ the probability distribution $\mu_{f;H}(\cdot \mid \sigma_{V_{\pinzero}}=\mathbf{0},\sigma_{V_{\pinone}}=\mathbf{1}, \bigcap_{W\in \mathcal{V}_{\eq}}\sigma_{W}^{\eq})$.
\end{definition}
 
\begin{remark}
Frequently, instead of formally specifying   $\mathcal{V}$, we will specify $\mathcal{V}$ implicitly by just saying ``consider the conditional distribution $\mu_{f;H}^{\condV}$ where the vertices in $V_{\pinzero}$ are pinned to 0, the vertices in  $V_{\pinone}$ are pinned to 1 and for all $W\in \mathcal{V}_{\eq}$, all the vertices in $W$ are forced to be equal".
\end{remark}

\subsection{Simulating  hard functions and inapproximability results}\label{sec:simulates}
 
We can now give a formal definition of ``simulation'', along the lines
that was informally discussed in Section~\ref{sec:sketcha} (Equation~\eqref{eq:examplesim}).
 
\begin{definition}\label{def:simulate}
Let $f:\{0, 1\}^k \ra \Rplus$ and $g:\{0, 1\}^t \ra \Rplus$. The function $f$ simulates the function $g$ if there 
is a $k$-tuple  hypergraph $H$, an admissible set $\mathcal{V}$ for $H$ with respect to $f$, and $t$ vertices $v_1, v_2, \ldots, v_t$ of~$H$ such that, for all $(s_1, s_2, \ldots, s_t) \in \{0, 1\}^t$,
\[\mu_{f;H}^{\condV}(\sigma(v_1) = s_1, \sigma(v_2) = s_2,\ldots,\sigma(v_t) = s_t) = \frac{g(s_1, s_2, \ldots, s_t)}{\sum\limits_{(s_1', s_2', \ldots, s_t') \in \{0, 1\}^t}g(s_1', s_2', \ldots, s_t')}.\]
If $\mathcal{V}=(\emptyset,\emptyset,\emptyset)$, then we say that $f$ \emph{perfectly simulates $g$}.
More generally, we say that $f$ simulates a set of functions $\mathcal{G}$ if $f$ simulates every $g\in \mathcal{G}$.
\end{definition}

The connection betweeen ``hard'' as defined in Definition~\ref{def:hard-function} and
intractability is given in the following lemma.     
     
\begin{lemma}[{\cite[Lemma 18]{symmetric}}]\label{lem:NPinapprox}
Let $f:\{0,1\}^k\rightarrow \Rplus$. If $f$ simulates a hard function, then for all sufficiently large $\Delta$, there exists $c>1$ such that $\HyperSpinf$ is $\NP$-hard.\qed
\end{lemma}

\begin{remark}
\cite[Lemma 18]{symmetric} is stated for symmetric functions, but the proof in \cite{symmetric} 
also works for asymmetric functions.  
\end{remark}

\section{Proof Sketch}\label{sec:sketchb} 

In this section, for a Boolean function $f:\{0, 1\}^k \ra \{0, 1\}$, we consider the complexity of the  problems 
$\HyperSpinfAP$
and 
$\HyperSpinf$.
Classifying the complexity of these problems  is the most important step in  the proof of Theorem~\ref{thm:main}. Namely, to obtain Theorem~\ref{thm:main}, it suffices to show that for every non-affine function $f$, we have that:
\begin{itemize}
\item If $f$ is in $\IM$, then for all sufficiently large $\Delta$, $\HyperSpinfAP$ is $\nBIS$-equivalent.  
\item If $f$ is not in $\IM$, then for all sufficiently large $\Delta$, there exists a real number $c>1$ such that $\HyperSpinf$ is $\NP$-hard.
\end{itemize}
Our main technical theorem to prove this is the following classification of Boolean functions, which  asserts that every non-affine function either supports perfect equality or simulates a hard function. 
\newcommand{\statethmasymmain}{Let $k\geq 2$ and let $f:\{0, 1\}^k \ra \{0, 1\}$ be a   Boolean function. Then at least one of three following propositions is true:
\begin{enumerate}
\item $f$ is affine;
\item $f$ supports perfect equality;
\item $f$ simulates a hard function.
\end{enumerate}}
\begin{theorem}\label{thm:asym-main}
\statethmasymmain{}
\end{theorem}
Theorem~\ref{thm:asym-main} is proved in Section~\ref{sec:asym-main}.
When $f$ simulates a hard function, using Lemma~\ref{lem:NPinapprox}, we can immediately conclude that for all sufficiently large $\Delta$, there exists $c>1$ such that $\HyperSpinf$ is $\NP$-hard. As we already discussed in Section~\ref{sec:sketcha}, it is important that, in the case where $f$ does not simulate a hard function, Theorem~\ref{thm:asym-main} guarantees that $f$ supports \emph{perfect} equality (rather than  simple imperfect equality); this allows us to recover the connection to  $\nBIS$ for those $f\in \IM$. In fact, when $f$ supports perfect equality, we can effectively carry out (a strengthening of) the program in \cite{trichotomy} to obtain the following classification which perfectly aligns with Theorem~\ref{thm:main}.
\newcommand{\statethmperfect}{Let $f:\{0,1\}^k\rightarrow \{0,1\}$ 
be a Boolean function that is not affine.  Suppose that $f$ supports perfect equality.
\begin{enumerate}
\item If $f$ is in $\IM$, then for all sufficiently large $\Delta$, $\HyperSpinfAP$ is $\nBIS$-equivalent.  
\item If $f$ is not in $\IM$, then for all sufficiently large $\Delta$, there exists a real number $c>1$ such that $\HyperSpinf$ is $\NP$-hard. 
\end{enumerate}}
\begin{theorem}\label{thm:perfect}
\statethmperfect{}
\end{theorem}
Theorem~\ref{thm:perfect} is proved in Section~\ref{sec:equalitysec}.
Thus, Theorems~\ref{thm:asym-main} and~\ref{thm:perfect} together achieve the desired classification of $\HyperSpinfAP$ when $f\in \IM$ as well as the  strong inapproximability results when $f\notin \IM$. Before delving into the proofs of Theorems~\ref{thm:asym-main} and~\ref{thm:perfect} however, it will be instructive to give the main ideas behind the proofs, especially of the more critical Theorem~\ref{thm:asym-main}.

To prove Theorem~\ref{thm:asym-main}, our proof departs from the previous approaches in the related works \cite{trichotomy} and \cite{symmetric}. In these works, $f$ was used to directly encode a binary function which was feasible because of the presence of equality in \cite{trichotomy} and the symmetry of $f$ in \cite{symmetric}. Instead, we take a much more painstaking combinatorial approach by using induction on the arity of the function $f$. 

The base case of the induction (proving Theorem~\ref{thm:asym-main} for arity-2 functions) is fairly simple to handle, so let us focus on the induction step. The rough idea, to put the induction hypothesis to work, is to study whether $f$ supports pinning-to-0 or pinning-to-1; then, provided that at least one these pinnings is available, we need to pin appropriately some arguments of $f$ to obtain a function $h$ of smaller arity. Our goal is then to ensure that $h$ is non-affine; then, we can invoke the induction hypothesis and obtain that $h$ either supports perfect equality or simulates a hard function. From there,  since $h$ was obtained by pinning some arguments of $f$, we will obtain by a transitivity argument (cf. Lemma~\ref{lem:reduction}) that $f$ either supports perfect equality or $f$ simulates the same hard function as $h$. (A detail here is that, in the case where $h$ supports perfect equality, to conclude that $f$ supports perfect equality from Lemma~\ref{lem:reduction}, we need to ensure that the pinnings of $f$ used to obtain $h$ were perfect.)

 Determining which arguments of $f$ need to be pinned is the most challenging aspect of this scheme. 
 Our  method for reducing the number of functions under consideration is to symmetrise $f$ in a natural way  and obtain a new function $f^*$ which is now symmetric (see Definition~\ref{def:fstar}). Then, it turns out that there are seven possibilities for the function $f^*$ which we need to  consider in detail (the functions are given in Definition~\ref{def:easyk}). That is, when the symmetrisation of $f$ is one of these seven functions, we have to figure out whether $f$ supports perfect equality and,  if not, work out the combinatorial structure of $f$ and pinpoint which arguments are suitable to be pinned. The details of the argument can be found in Section~\ref{sec:bbb}.

The proof of Theorem~\ref{thm:perfect}, where $f$ supports perfect equality, basically follows the approach of \cite{trichotomy}. However, to get the stronger inapproximability results, we have to take a detour studying self-dual functions (functions whose value does not change when we complement their arguments).
We show that if $f$ is self-dual then it simulates a hard function (Theorem~\ref{thm:sd-main}). The problem with self-dual functions is that they do not support pinning-to-0 or pinning-to-1,  so we are not able to  use the relevant results from~\cite{trichotomy}. After proving Theorem~\ref{thm:sd-main} and doing some preparatory work in Section~\ref{sec:aaa} to ensure that ``implementations in CSPs" work in the repeat-free setting when $f$ supports perfect equality (see Lemma~\ref{lem:transitivity}), the techniques of \cite{trichotomy} can be adapted to get Theorem~\ref{thm:perfect}.

\section{Notation and results from the literature}
\subsection{Notation}\label{sec:notation}

For a vector $\xb$, we use $x_i$ to denote the $i$'th entry of $\xb$. 
Further, for vectors $\xb$ and $\yb$ of the same  length, $\xb\oplus \yb$ will denote the coordinate-wise addition of $\xb$ and $\yb$ over $\mathrm{GF}(2)$. More generally, for any binary Boolean operator $\otimes$, we will denote by $\xb\otimes\yb$ the vector whose $i$-th entry is given by $x_i\otimes y_i$. We will use $\zeroes,\ones$ to denote the vectors whose entries are all zeros and all ones, respectively (the  length of these vectors will be clear from context). Finally, for a Boolean vector $\xb$, $\overline{\xb}$ will denote the coordinate-wise ``negation" of $\xb$, i.e., $\overline{\xb}=\xb\oplus \ones$. For a positive integer~$k$, $[k]$ denotes $\{1,\ldots,k\}$.

\begin{definition}[$\Omega_f$, $\chi_S$]
Let $f: \{0, 1\}^k \rightarrow \Rplus$. 
For $S\subseteq [k]$,
$\chi_S$ denotes the  
characteristic vector of~$S$, which is the
 length-$k$ Boolean vector such that, for all $i\in[k]$, 
the $i$-th bit of $\chi_S$ is $1$ if and only if $i \in S$.
Finally, $\Omega_f = \{S \subseteq [k] \mid \chi_S \in R_f\}$, where $R_f$ is the relation underlying~$f$, defined at the
beginning of Section~\ref{sec:prelim}.
\end{definition}

\begin{definition}[The symmetrisation $f^*$]\label{def:fstar}
Let $f: \{0, 1\}^k  \rightarrow \Rplus$. We denote by $f^*$ the symmetrisation of $f$ obtained as follows. Let $S_k$ denote the set of all permutations $\pi:[k]\rightarrow [k]$. Then $f^*:\{0,1\}^k \rightarrow \Rplus$ is the function defined  by
\[f^*(x_1,\hdots,x_k)=\prod_{\pi\in S_k} f(x_{\pi(1)},\hdots,x_{\pi(k)}).\] \end{definition}

\subsection{Affine functions}
The following well-known characterisation of affine functions (cf. Definition~\ref{def:affine}) is instructive and will be useful later.
For a proof, see, for example, Lemma~4.10 of~\cite{CKS} or Lemma~11 of~\cite{trichotomy}.

\begin{lemma} \label{lem:affine}
Let $f:\{0,1\}^k\ra\{0,1\}$ be a Boolean function. Then:
\begin{enumerate}
\item \label{it:abc} $f$ is affine iff for every $\ab,\bb,\cb\in R_f$, it holds that $\ab\oplus\bb\oplus \cb\in R_f$.
\item \label{it:fixeda} If $f$ is not affine, then for every $\ab\in R_f$, there  are $\bb,\cb\in R_f$ such that  
$\ab\oplus\bb\oplus \cb\notin R_f$.\qed
\end{enumerate}
\end{lemma}

The set of affine \emph{symmetric} Boolean functions $f:\{0,1\}^k\rightarrow\{0,1\}$ is given by the following set $\EASYk$.
\begin{definition}\label{def:easyk}
For $k\geq 2$, let $\EASYk$ be the set
containing the following seven functions.
\begin{align*}
\zerof^{(k)}(x_1,\ldots,x_k)&=0,\quad
\onef^{(k)}(x_1,\ldots,x_k)=1,\quad 
\allzerof^{(k)}(x_1,\ldots,x_k)=\mathbf{1}\{x_1=\hdots=x_k=0\},\\
\allonef^{(k)}(x_1,\ldots,x_k)&=\mathbf{1}\{x_1=\hdots=x_k=1\},\quad 
\eqf^{(k)}(x_1,\ldots,x_k)=\mathbf{1}\{x_1=\hdots=x_k\}, \\
\evenf^{(k)}(x_1,\ldots,x_k)&=\mathbf{1}\{x_1\oplus\cdots\oplus x_k=0\},  \quad
\oddf^{(k)}(x_1,\ldots,x_k)=\mathbf{1}\{x_1\oplus\cdots\oplus x_k=1\}.
\end{align*}
\end{definition}

\subsection{A characterisation of $\IM$}
In the language of universal algebra,
Creignou, Kolaitis, and Zanuttini~\cite{CKS} 
have shown that $\IM$ (see Definition~\ref{def:IM2}) is precisely the ``co-clone'' corresponding to 
the ``clone''~$M_2$ in   Post's lattice (see~\cite{Post}).
Defining clones and co-clones would be a bit of a distraction from this paper, but
the only fact that we need is the following (which follows directly from the
definitions of clones and co-clones and from the fact that $\IM$ is the co-clone corresponding to
$M_2$).

\begin{lemma}\label{lem:IM2}
Let $f:\{0,1\}^k\ra\{0,1\}$ be a Boolean function. Then, the function $f$ is in $\IM$ iff for every $\xb,\yb\in R_f$ it holds that $\xb\vee \yb \in R_f$ and $\xb\wedge \yb \in R_f$.\qed
\end{lemma}

\subsection{The case where $f$ is symmetric: extensions to the asymmetric case}
In this section, we state a few results from \cite{symmetric} which were stated for the case where $f$ is a symmetric Boolean function, but whose proof works just as well even when $f$ is asymmetric. 

The following lemma, which is Lemma~12 of~\cite{symmetric}, 
gives sufficient conditions for pinning-to-0, pinning-to-1 and equality. 
The statement of the lemma in~\cite{symmetric} is restricted to symmetric functions~$f$, but
the proof applies to all functions (with the trivial modification that the 
vertices in the hyperarcs in the constructed
 $k$-tuple hypergraph~$H$ must be  ordered appropriately).
\begin{lemma}[{\cite[Lemma 12]{symmetric}}]\label{lem:gadgets}
Let $f:\{0,1\}^k\rightarrow \Rplus$ and let
$H$ be a  $k$-tuple hypergraph.
\begin{enumerate}
\item \label{it:pinning0} If there  is a vertex $v$ in $H$ such that $\mu_{f;H}(\sigma_v=0)>\mu_{f;H}(\sigma_v=1)$, then $f$ supports pinning-to-0.
\item \label{it:pinning1} If there is a vertex $v$ in $H$ such that $\mu_{f;H}(\sigma_v=1)>\mu_{f;H}(\sigma_v=0)$, then $f$ supports pinning-to-1.
\item \label{it:equality} If there  are  vertices $x,y$ in $H$ such that $\mu_{f;H}(\sigma_{x}=\sigma_{y}=0)=\mu_{f;H}(\sigma_{x}=\sigma_{y}=1)$ and $\mu_{f;H}(\sigma_{x}=\sigma_{y})>\mu_{f;H}(\sigma_{x}\neq \sigma_{y})$, then $f$ supports equality.\qed
\end{enumerate}
\end{lemma}

\begin{lemma}[{\cite[Lemma 17]{symmetric}}]\label{lem:generalpinfull}
Let $f:\{0,1\}^k\rightarrow\Rplus$. Let $H=(V,\calF)$ be a  $k$-tuple hypergraph  and let $S$ be a subset of~$V$. Let $\mathcal{V}$ be  admissible  for~$H$ with respect to~$f$. 
Then, for every $\epsilon>0$, there is a 
$k$-tuple
hypergraph~$H'=(V',\mathcal{F}')$  with  $V\subseteq V'$ and $\mathcal{F} \subseteq \mathcal{F}'$ such that, for every $\tau:S\rightarrow \{0,1\}$, it holds that 
\[\big|\mu_{f;H'}(\sigma_S=\tau)-\mu_{f;H}^{\condV}(\sigma_S=\tau)\big|\leq \epsilon,\]
where $\mu_{f;H}^{\condV}(\cdot)$ is as in Definition~\ref{def:generalpinfull}.\qed
\end{lemma}

We will also use the following result from \cite{symmetric} which applies to \emph{symmetric} Boolean functions. 
\begin{lemma}[{\cite[Proof of  Theorem 3]{symmetric}}]\label{cor:corfsymm}
Let $k\geq 2$ and let $f:\{0,1\}^k\rightarrow\{0,1\}$ be a \emph{symmetric} Boolean function such that $f\notin\mathsf{EASY}(k)$. Then either $f$  simulates a hard function or $f$ supports perfect equality (or both).
\end{lemma}
\begin{proof}
We briefly overview the proof in \cite{symmetric}, the relevant parts are in \cite[Section 4]{symmetric}. 
\begin{enumerate}
\item \cite[Lemma 13]{symmetric} shows that every function $f\notin\mathsf{EASY}(k)$ supports one of pinning-to-0, pinning-to-1 or equality.
\item In \cite[Section 4.1]{symmetric}, the case where $f$ supports both pinning-to-0 and pinning-to-1 is considered. Then, \cite{symmetric} shows that $f$ simulates a hard function. 
\item In \cite[Section 4.2]{symmetric}, the case where $f$ supports equality but neither pinning-to-0 nor pinning-to-1 is considered. The proof splits into cases depending on whether $f(\zeroes)=0$ or $f(\zeroes)=1$. When $f(\zeroes)=0$ (\cite[Section 4.2.2]{symmetric}), \cite{symmetric} shows that $f$ supports perfect equality (\cite[Lemma 28]{symmetric}). When $f(\zeroes)=1$ (\cite[Section 4.2.3]{symmetric}), \cite{symmetric} shows that $f$ simulates a hard function.
\item  In \cite[Section 4.3]{symmetric}, the case where $f$ supports pinning-to-0 is considered. Then, \cite{symmetric} shows that $f$ simulates a hard function. (The case where  $f$ supports pinning-to-1 is identical by switching the spins 0 and 1.)
\end{enumerate}
Thus, for every symmetric function $f:\{0,1\}^k\rightarrow\{0,1\}$ such that $f\notin\mathsf{EASY}(k)$, the results of \cite{symmetric} show that either $f$  simulates a hard function or $f$ supports perfect equality.
\end{proof}

\section{Non-affine Boolean functions either support perfect equality or simulate a hard function}\label{sec:asym-main}
In this section, we prove Theorem~\ref{thm:asym-main}, i.e.,  that every non-affine Boolean function either supports perfect equality or simulates a hard function. Before proving the theorem, we will need a few technical lemmas.

\subsection{A few preparatory lemmas}
\begin{lemma}\label{lem:allonepinning}
Let $f:\{0, 1\}^k \ra \{0, 1\}$ be a $k$-ary Boolean function. Suppose that $f^*= f_\mathsf{allone}$. Then $f$ supports perfect pinning-to-$1$.
\end{lemma}

\begin{proof}
Let $H = (V, \mathcal{F})$ be the  $k$-tuple hypergraph with $V = \{v_1, v_2, \ldots, v_k\}$ and $\mathcal{F} = \{e_\pi\mid \pi \in S_k\}$ where $e_\pi = (v_{\pi(1)}, v_{\pi(2)}, \ldots, v_{\pi(k)})$. Since $f^*= f_\mathsf{allone}$, we have that for all $\sigma: V \ra \{0, 1\}$ it holds that  $w_{f, H}(\sigma) > 0$ if and only if $\sigma(v_1) = \sigma(v_2) = \cdots = \sigma(v_k) = 1$. Thus, $f$ supports perfect pinning-to-$1$.
\end{proof}

Completely analogously, we have the following pinning lemma when $f^*=f_\mathsf{allzero}$.
\begin{lemma}\label{lem:allzeropinning}
    Let $f:\{0, 1\}^k \ra \{0, 1\}$ be a $k$-ary Boolean function. Suppose that $f^*= f_\mathsf{allzero}$. Then $f$ supports perfect pinning-to-$0$.\qed
\end{lemma}

For any function $f$ such that $f^*= f_{\mathsf{zero}}$, we have the following pinning lemma.
\begin{lemma}\label{lem:zeropinning}
Let $f:\{0, 1\}^k \ra \{0, 1\}$ be a $k$-ary Boolean function. Suppose that $f\neq f_\mathsf{zero}$ and $f^*= f_\mathsf{zero}$. Then at least one of the following two propositions is true:
\begin{enumerate} 
\item  $f$ supports perfect pinning-to-$0$ and perfect pinning-to-$1$;
\item $f$ supports perfect equality.
\end{enumerate}
\end{lemma}
\begin{proof} 
Note that the conditions in the lemma imply that $k\geq 2$.
Let $S_k$ denote the set of all permutations $\pi:[k]\rightarrow [k]$ and 
let $\mathrm{id} \in S_k$ denote the identity permutation. For any subset $A\subseteq S_k$,
let $f_A$ be the function defined by   $f_{A}(w_1,\hdots,w_k):=\prod_{\pi\in A}f(w_{\pi(1)},\hdots,w_{\pi(k)})$. 
Note that $f_{S_k}=f^*= f_\mathsf{zero}$.
Also,   for any $\pi\in S_k$ we have   $f_{\{\pi\}}\neq f_\mathsf{zero}$ (since $f\neq f_\mathsf{zero}$). 
By iteratively removing permutations from $S_k$  we will thus obtain a   subset $T\subseteq S_k$ with $|T|>1$ such that  $f_{T}=f_\mathsf{zero}$ and, for every $\pi\in T$, it holds that $f_{T\backslash \{\pi\}}\neq f_\mathsf{zero}$. By renaming the variables if necessary, we may assume that $\mathrm{id}\in T$.

Let $H_0=(V_0,\mathcal{F}_0)$ be the  $k$-tuple hypergraph with vertex set $V_0=\{x_1,\hdots,x_k\}$ and hyperarc set $\mathcal{F}_0=\cup_{\pi\in T\backslash \{\mathrm{id}\}}\{(x_{\pi(1)},\hdots,x_{\pi(k)})\}$. By the choice of the set $T$, we have that $Z_{f;H_0}>0$. For $i=1,\hdots,k$, let $H_i=(V_i,\mathcal{F}_i)$ be the  $k$-tuple hypergraph with vertex set $V_{i}=V_0\cup \{y_{i+1},\hdots,y_k\}$ and hyperarc set $\mathcal{F}_i=\mathcal{F}_0\cup \{(x_1,\hdots,x_i,y_{i+1},\hdots, y_k)\}$. Again by the choice of the set $T$ we have that $Z_{f;H_k}=0$. Thus, there exists $0\leq j<k$ such that $Z_{f;H_j}>0$ and $Z_{f;H_{j+1}}=0$. Note that for every assignment $\sigma:V_j\rightarrow \{0,1\}$ with $w_{f; H_j}(\sigma)>0$ it holds that $\sigma(x_{j+1})\neq \sigma(y_{j+1})$; otherwise, for the assignment $\sigma'=\sigma_{V_{j+1}}$ (i.e., the restriction of the assignment $\sigma$ to the set $V_{j+1}$), it would hold that $w_{f; H_{j+1}}(\sigma')>0$, contradicting that $Z_{f;H_{j+1}}=0$. 

For $s_1,s_2\in\{0,1\}$, let
\begin{equation*}
Z_{s_1,s_2}:=\sum_{\substack{\sigma:V_{j}\rightarrow \{0,1\};\\ \sigma(x_{j+1})=s_1,\, \sigma(y_{j+1})=s_2}}w_{f; H_{j}}(\sigma)
\end{equation*}
By the argument above, we have that $Z_{00}=Z_{11}=0$. Since $Z_{f;H_{j}}>0$, we have that at least one of $Z_{01}$ and $Z_{10}$ is non-zero. In fact, we may assume that both are non-zero, since otherwise $f$ supports 
both 
perfect  pinning-to-0 and perfect pinning-to-1 
so proposition~1 in the statement of the lemma is satisfied
(for example, if $Z_{10}=0$, then $\mu_{f;H_{j}}(\sigma(x_{j+1})=0)=1$ and $\mu_{f;H_{j}}(\sigma(y_{j+1})=1)=1$).

Let $J_1,J_2$ be two disjoint copies of $H_{j}$. Denote by $u_1,u_2$ the vertices corresponding to $x_{j+1}$  in $J_1,J_2$, respectively. Also, denote by $v_1,v_2$ the vertices corresponding to $y_{j+1}$ in $J_1,J_2$. Let $J=(V,\mathcal{F})$ be the $k$-tuple hypergraph obtained by taking the union of $J_1$ and $J_2$ and identifying the vertices $u_2$ and $v_1$ into a new vertex $w$ (i.e., we merge the vertex corresponding to $x_{j+1}$ in $J_2$ and  the vertex corresponding to $y_{j+1}$ in $J_1$).

For $s_1,s_2\in\{0,1\}$, let
\begin{equation*}
Z_{s_1,s_2}':=\sum_{\substack{\sigma:V\rightarrow \{0,1\};\\ \sigma(u_1)=s_1,\, \sigma(v_2)=s_2}}w_{f; J}(\sigma).
\end{equation*}
By considering the spin of the vertex $w$, we obtain that
\[Z_{s_1,s_2}'=Z_{s_1,0}Z_{0,s_2}+Z_{s_1,1}Z_{1,s_2},\]
which gives that
\[Z_{00}'=Z_{01}Z_{10},\quad Z_{01}'=0, \quad Z_{10}'=0,\quad Z_{11}'=Z_{10}Z_{01}.\]
Since $Z_{01},Z_{10}\neq 0$, we obtain  that 
\[\mu_{f;J}(\sigma(u_1)=\sigma(v_2)=0)=\mu_{f;J}(\sigma(u_1)=\sigma(v_2)=1)=\frac{1}{2},\]
and hence $f$ supports perfect equality.
\end{proof}

For any function $f$, we can show that if $f^*$ is  $f_{\mathsf{EQ}}$, $f_{\mathsf{odd}}$ or $f_{\mathsf{even}}$ then $f$ supports perfect equality.

\begin{lemma}\label{lem:equalizer}
Let $f:\{0, 1\}^k \ra \{0, 1\}$ be a $k$-ary Boolean function. Suppose that $f^* = f_{\mathsf{EQ}}$. Then $f$ supports  perfect equality.
\end{lemma}

\begin{proof}
Let $H = (V, \mathcal{F})$ be the $k$-tuple hypergraph with $V = \{v_1, v_2, \ldots, v_k\}$ and hyperarc set $\mathcal{F} = \{e_\pi\mid \pi \in S_k\}$ where $e_\pi = (v_{\pi(1)}, v_{\pi(2)}, \ldots, v_{\pi(k)})$. Since $f^*= f_{\mathsf{EQ}}$, we have that for all $\sigma: V \ra \{0, 1\}$ it holds that  $w_{f, H}(\sigma) > 0$ iff $\sigma(v_1) = \sigma(v_2) = \cdots = \sigma(v_k) = 1$ or $\sigma(v_1) = \sigma(v_2) = \cdots = \sigma(v_k) = 0$. Thus, $f$ supports  perfect equality.
\end{proof}

\begin{lemma}\label{lem:parity}
Let $f:\{0, 1\}^k \ra \{0, 1\}$ be a $k$-ary Boolean function. Suppose that $f^* \in \{ f_{\mathsf{odd}},f_{\mathsf{even}}\}$. Then $f$ supports perfect equality.
\end{lemma}

\begin{proof}
Let $H = (V, \mathcal{F})$ be the $k$-tuple hypergraph with $V = \{v_1, v_2, \ldots, v_{k + 1}\}$ and $\mathcal{F} = \{e_\pi\mid \pi \in S_k\}\cup \{e'_\pi\mid \pi\in S_k\}$ where $e_\pi = (v_{\pi(1)}, v_{\pi(2)}, \ldots, v_{\pi(k)})$ and $e'_\pi = (v_{\pi(1) + 1}, v_{\pi(2) + 1}, \ldots, v_{\pi(k)+1})$ (note that $H$ has $k+1$ vertices and $2k!$ hyperarcs).  Since $f^*$
is either $f_{\mathsf{odd}}$ or $f_{\mathsf{even}}$, for all $\sigma: V \ra \{0, 1\}$ with $w_{f, H}(\sigma) > 0$, we have that the parity of number of ones among $\sigma(v_1), \sigma(v_2), \ldots, \sigma(v_k)$ and the parity of number of ones among $\sigma(v_2), \sigma(v_3), \ldots, \sigma(v_{k+1})$ must be the same and thus $\sigma(v_1) = \sigma(v_{k+1})$. Furthermore, for $s\in \{0,1\}$, there are exactly $2^{k-1}$ assignments $\sigma:V\rightarrow \{0,1\}$ such that $w_{f;H}(\sigma)>0$, $\sigma(v_1) = \sigma(v_{k+1})$ and $\sigma(v_2) \oplus \sigma(v_3) \oplus \cdots \oplus \sigma(v_k) = s$.  It follows that 
\[\mu_{f;H}(\sigma(v_1) = \sigma(v_{k+1}) = 0) = \mu_{f;H}(\sigma(v_1) = \sigma(v_{k+1}) = 1) = \frac{1}{2},\] 
which means that $f$ supports perfect equality.
\end{proof}

By the above lemmas, we can show that some functions can be either dealt with directly, or reduced to other functions with smaller arity.

\begin{definition}\label{def:dopin}
For $s\in\{0,1\}$, 
let $\delta_s: \{0,1\} \rightarrow \{0,1\}$ be the Boolean function defined by
$\delta_s(s)=1$ and $\delta_s(1 \oplus s)=0$.
Define $f_{i\ra s}$ to be the function obtained from $f$ by pinning the $i$-th argument of $f$ to $s$, i.e. 
 \[
 f_{i\ra s}(x_1, \ldots,x_{i-1},x_{i+1},\ldots,x_k) = \sum_{x_i\in\{0,1\}} f(x_1, \ldots, x_k) \cdot \delta_s(x_i).\] 
Similarly, for disjoint $S,T\subseteq [k]$, let $f_{S\ra 0, T\ra 1}$ be the $(k-\abs{S \cup T})$-ary function obtained from $f$ by pinning the arguments in $S$ to 0 and the arguments in $T$  to 1
So if ${\mathbf{x}}'$ denotes the $|S\cup T|$-ary vector containing all $x_i$ with $i\in S\cup T$
and ${\mathbf{x}}''$ denotes the $k-|S \cup T|$-ary vector containing all   $x_i$ with $i\in [k]\setminus S \cup T$,
\[f_{S\ra 0, T \ra 1}({\mathbf {x}}'') = \sum_{ {\mathbf {x}}'\in \{0,1\}^{|S\cup T|}} f(x_1, \ldots, x_k) \cdot \prod_{i\in S} \delta_0(x_i) \cdot \prod_{j \in T} \delta_1(x_j).\]
If $S = \emptyset$ or $T = \emptyset$, we will omit $S \ra 0$ or $T \ra 1$ from the notation.
\end{definition}

Using  Definition~\ref{def:dopin}, we have the following lemma:

\begin{lemma}\label{lem:reduction}
Let $f:\{0, 1\}^k \ra \{0, 1\}$ be a Boolean function. Suppose that $S_0$ and $S_1$ are disjoint subsets of $[k]$ such that, for $a\in\{0,1\}$, $S_a$ is empty if $f$ does \emph{not} support perfect pinning-to-$a$. Let $h = f_{S_0\ra 0,S_1\ra 1}$.
\begin{enumerate}
\item \label{it:4335rge} \label{item:oldone}
If $h$ supports equality, then $f$ also supports equality. Further, if $h$ supports perfect equality, then $f$ also supports perfect equality.
\item \label{it:5643rgb} \label{item:oldtwo}
If $h$ supports pinning-to-$s$ for some $s \in \{0, 1\}$, then $f$ also supports pinning-to-$s$.
\item\label{item:newthree} 
If $h$ simulates a function  $g:\{0, 1\}^2 \ra \Rplus$
that is not~$\zerof^{(2)}$
then $f$ simulates $g$ as well. Also, if $h$ perfectly simulates $g$ then $f$  perfectly simulates $g$ as well. 
\end{enumerate}
\end{lemma}

\begin{proof}
Without loss of generality, we assume that the arity of $h$ is $n$, and that $S_0\cup S_1 = \{n + 1, n + 2, \ldots, k\}$.
For each $a \in \{0,1\}$, if $S_a$ is non-empty, then by assumption $f$ supports perfect pinning-to-$a$, so there exists a $k$-tuple hypergraph $H_a=(V_a,\mathcal{F}_a)$ with a vertex $w_a\in V_a$ such that $\mu_{f;H_a}(\sigma_{w_a} = a)=1$. 

We now give a general construction 
which takes any $n$-tuple hypergraph~$H = (V, \mathcal{F})$ 
and produces a new $k$-tuple hypergraph~$H'=(V',\mathcal{F}')$.
To do this, we take $k-n$ new vertices $v'_{n+1},\ldots,v'_k$ that are not in~$V$
and let $V' = V \cup \{v'_{n+1},\ldots,v'_k\}$.
The hyperarcs of~$H'$ are in one-to-one correspondence with those in~$H$:
For each hyperarc $(u_1, u_2, \ldots, u_n)$ in $H$, 
we add the hyperarc $(u_1, u_2, \ldots, u_n, v'_{n + 1}, v'_{n + 2}, \ldots, v'_k)$ to $H'$. 
Moreover,  for $i\in S_0$, add a distinct copy of $H_0$ to $H'$  by identifying $v'_i$ with the vertex $w_0$ in $H_0$. Also, for $i\in S_1$, add a distinct copy of $H_1$ to $H'$  by identifying $v'_i$ with the vertex $w_1$ in $H_1$.  
  
Say that an assignment   
$\sigma:V'\rightarrow \{0,1\}$
is \emph{relevant} if, for each
$a\in\{0,1\}$ and each $i\in S_a$, $\sigma(v_i) =a$.
The copies of $H_0$ and $H_1$ ensure that, for every 
assignment $\sigma:V'\rightarrow \{0,1\}$  with $w_{f;H'}(\sigma)>0$,
$\sigma$ is relevant. 
The definition of $h$ ensures that, for any relevant assignment~$\sigma$,
 \begin{equation}\label{eq:matchH}
 w_{f;H'}(\sigma)=w_{h;H}(\sigma_V).
 \end{equation}  
 We now use~\eqref{eq:matchH} to establish the three items in the statement of the lemma.

\begin{enumerate}

\item Suppose that $h$ supports equality. For any $\epsilon \in (0,1)$, there is  an $n$-tuple hypergraph $H = (V, \mathcal{F})$ and two vertices $x$ and $y$ of $H$ such that, for every $s \in \{0, 1\}$, $\mu_{h;H}(\sigma_x = \sigma_y = s) \geq (1 - \epsilon)/2$.  
Construct $H'$ from~$H$ using the general construction above.
From~\eqref{eq:matchH}, we conclude that,
  for any $s\in \{0,1\}$, 
 $\mu_{f;H'}(\sigma_x = \sigma_y = s)=\mu_{h;H}(\sigma_x = \sigma_y = s)\geq (1-\epsilon)/2$, 
so $f$ supports equality.
If $h$ supports \emph{perfect} equality, then we can take $\epsilon=0$ in this argument,
obtaining the conclusion that 
   $f$ also supports \emph{perfect} equality.

\item Suppose that $h$ supports pinning-to-$s$.
For any  $\epsilon \in (0, 1)$ there is an $n$-tuple hypergraph 
$H = (V, \mathcal{F})$ and a vertex $x$ of $H$ such that $\mu_{h;H}(\sigma_x = s) \geq 1 - \epsilon$. 
Construct $H'$ from~$H$ using the general construction above.
From~\eqref{eq:matchH}, we conclude that,
   $\mu_{f;H'}(\sigma_x =   s)=\mu_{h;H}(\sigma_x =  s)\geq  1-\epsilon$, 
so $f$ supports  pinning-to-$s$.
 
 \item
Let  $g:\{0, 1\}^2 \ra \Rplus$ be a function
that is not~$\zerof^{(2)}$.
 Suppose first that $h$   simulates $g$. 
 By the definition of ``simulates", there exists an $n$-tuple hypergraph $H$ with admissible $\mathcal{V}$ (with respect to $h$) and two vertices $u$ and $v$ in $H$ such that, for every $s, t\in \{0, 1\}$, it holds that
\begin{equation}\label{eq:56g5tg5}
\mu_{h;H}^{\condV}(\sigma(u) = s, \sigma(v) = t) = \frac{g(s, t)}{\sum_{i, j\in \{0, 1\}}g(i, j)}. \end{equation}
Since $g\neq \zerof^{(2)}$, the expression in~\eqref{eq:56g5tg5} is well-defined.

Construct $H'$ from~$H$ using the general construction above.
From Items~1 and~2 of the lemma,
if $h$ supports equality or pinning-to-$0$ or pinning-to-$1$ then so does~$f$.
Thus, $\mathcal{V}$ is  admissible for $H'$ with respect to $f$. 
It follows from~\eqref{eq:matchH} that 
\begin{align*}
\mu^{\condV}_{h;H}(\sigma(u) = s, \sigma(v) = t) &= \mu_{f;H'}^{\condV}(\sigma(u) = s, \sigma(v) = t \mid\, \land_{i \in S_0} \sigma(v_i) = 0,\,\land_{i \in S_1} \sigma(v_i) = 1)\\
&=\mu_{f;H'}^{\condV}(\sigma(u) = s, \sigma(v) = t),
\end{align*}
so, using \eqref{eq:56g5tg5}, we obtain that $f$ simulates $g$, as wanted.
If $h$ \emph{perfectly} simulates $g$ then 
we can take $\mathcal{V} = (\emptyset,\emptyset,\emptyset)$, so the argument shows that $f$ perfectly simulates~$g$.

\end{enumerate}

\end{proof}

\subsection{Proof that every
non-affine Boolean functions either supports perfect equality or simulates a hard function}\label{sec:bbb}

\begin{definition}\label{def:semitrivial}
A function $f:\{0,1\}^k\rightarrow \{0,1\}$ is \emph{semi-trivial} if and only if there exists a set $S\subseteq[k]$ such that $\Omega_f = \{T \mid S \subseteq T \subseteq [k]\}$ or $\Omega_f = \{T \mid T\subseteq S\}$.  
\end{definition}

\begin{remark}
Every semi-trivial function $f$ is affine since $R_f$ equals the solution set of the system of equations of the form $\{x_i=1\}_{i\in S}$ or $\{x_i=0\}_{i\in S}$  where $S\subseteq[k]$ is as in Definition~\ref{def:semitrivial}.
\end{remark}

\begin{lemma}\label{lem:upclose}
Let $k\geq 2$ and let
$f:\{0, 1\}^k \ra \{0, 1\}$ be a $k$-ary Boolean function. Suppose that $f\neq f_\mathsf{allone}$ and $f^*= f_\mathsf{allone}$.
    Let $S$ be a set in $\Omega_f$  such that $S \neq [k]$. Then at least one of the following propositions is true:
    \begin{enumerate} 
        \item $\forall\, T\supseteq S$, we have $T \in \Omega_f$;
        \item $f$ supports perfect equality;
        \item $f$ simulates a hard function.
    \end{enumerate}
\end{lemma}

\begin{proof} Without loss of generality (by re-numbering the variables), let $S = \{n + 1, n + 2, \ldots, k\}$ for some integer $n \geq 1$.
By Lemma~\ref{lem:allonepinning}, $f$ supports perfect pinning-to-$1$. Let $h(x_1, x_2, \ldots, x_n) = f(x_1, \ldots, x_k)_{S\ra1}$. Note that $h^*(\zeroes)=1$ and $h^*(\ones)=1$. We may assume that $n\geq 2$ (otherwise $\forall\, T\supseteq S$, we have $T \in \Omega_f$).
\begin{itemize}
\item[Case]$1$.  $h^* \not\in  \mathsf{EASY}(n)$. In this case,
Lemma~\ref{cor:corfsymm} ensures that either $h$ simulates a hard function or $h$ supports perfect equality (or both).
If $h$ simulates a hard function, then
by Item~3 of Lemma~\ref{lem:reduction},  
$f$ also simulates a hard function.
If $h$ supports perfect equality, then, by 
Item~1 of Lemma~\ref{lem:reduction}, $f$ also supports perfect equality.
\item[Case]$2$.  $h^* \in \mathsf{EASY}(n)$. Then $h^* \in \{f_{\mathsf{one}},f_\mathsf{even}, f_\mathsf{EQ}\}$ since $h^*(\zeroes)=1$ and $h^*(\ones)=1$.
If $h^* = f_\mathsf{one}$, we have that $h(\xb) = 1$ for all $\xb \in  \{0, 1\}^n$. Since $h(\xb) = f_{S\ra1}$, we obtain that $T \in \Omega_f$ for all $T \supseteq S$.
    
If $h^* \in \{f_\mathsf{even}, f_\mathsf{EQ}\}$, then $h$ supports perfect equality by Lemmas~\ref{lem:equalizer} and~\ref{lem:parity}. Since $f$ supports perfect pinning-to-$1$, by 
Item~1 of Lemma~\ref{lem:reduction} we obtain that $f$ supports perfect equality as well.
\end{itemize}
This concludes the proof.
\end{proof}

\begin{lemma}\label{lem:allone}
Let $k\geq 2$ and let
$f:\{0, 1\}^k \ra \{0, 1\}$ be a $k$-ary Boolean function. Suppose that $f\neq f_\mathsf{allone}$ and $f^*= f_\mathsf{allone}$. Then at least one of the four following propositions is true:
\begin{enumerate} 
\item $f$ is semi-trivial;
\item there exists $t \in [k]$ such that $f_{t\ra1}$ is not affine;
\item \label{it:feqallone} $f$ supports perfect equality;
\item \label{it:hardallone} $f$ simulates a hard function.
\end{enumerate}
\end{lemma}

\begin{proof}
If $k = 2$, since $f\neq f_\mathsf{allone}$ and $f^*= f_\mathsf{allone}$, we have $f(1,1) = 1$, $f(0,0) = 0$ and exactly one of $f(0,1)$ and $f(1,0)$ is one, so $f$ is semi-trivial. Thus, for the rest of the proof we may assume that $k \geq 3$.
    
Since $f^*= f_\mathsf{allone}$, we have that $f(\ones)=1$ and $f(\zeroes)=0$. Further, since $f \neq f_{\mathsf{allone}}$, there exists $S\in \Omega_f$ with $\abs{S} < k$.
\begin{itemize}
\item[Case]$1$. Every $S \in \Omega_f$ satisfies $\abs{S} \geq k - 1$.
        
If there is only one set $S$ in $\Omega_f$ with $\abs{S} = k-1$, then we have that $f$ is semi-trivial (since $f(\ones)=1$). Otherwise, there are distinct sets $S, S' \in \Omega_f$ with $\abs{S} = \abs{S'} = k - 1$, so $\abs{S \cap S'} = k -2$ and thus $S \cap S' \neq\emptyset$ and $S \cap S'\not\in \Omega_f$. Let $t \in S \cap S'$. We claim that $h=f_{t\ra1}$ is not affine; to see this, note that $f(\chi_{S}) = f(\chi_{S'}) = f(\chi_{[k]}) = 1$ and $f(\chi_S \oplus \chi_{S'} \oplus \chi_{[k]}) = f(\chi_{S\cap S'}) = 0$. Since $h=f_{t\ra1}$ and $t\in S\cap S'$, we obtain that 
\[S\backslash \{t\},\, S'\backslash \{t\},\, [k]\backslash \{t\}\in \Omega_h \mbox{ but } (S\cap S')\backslash \{t\}\notin \Omega_h.\] 
By Item~\ref{it:abc} of Lemma~\ref{lem:affine}, it thus follows that $h$ is not affine, as wanted.
        
\item[Case]$2$. There exists $S \in \Omega_f$ with $\abs{S} \leq k - 2$.

Let $S$ be a set in $\Omega_f$ with the smallest cardinality among the sets in $\Omega_f$. By Lemma~\ref{lem:upclose}, either $f$ satisfies 
proposition~\ref{it:feqallone} or~\ref{it:hardallone},
in the statement of the lemma (so we are finished), 
 or every $Q \supseteq S$ satisfies $Q \in \Omega_f$. Thus, for the rest of the proof we may assume that for every $Q \supseteq S$ it holds that $Q \in \Omega_f$.
        
Let $\Psi = \{ W \in \Omega_f \mid S \setminus W \neq \emptyset\}$.        
If $\Psi$ is empty then $f$ is semi-trivial, so it satisfies proposition~1 in the statement of the lemma (and we are finished).
So assume that $\Psi$ is non-empty and choose $T\in \Psi$ with
cardinality as small as possible.

By the choice of $S$, $T$ cannot be a strict subset of $S$, so $T\setminus S$ is not empty.  Applying Lemma~\ref{lem:upclose} to the set $T$, we may assume that $\forall \,Q \supseteq T$ it holds that $Q \in \Omega_f$ (otherwise, $f$ will satisfy 
proposition~\ref{it:feqallone} or~\ref{it:hardallone},
in the statement of the lemma, so we are finished). Since $f(\zeroes) = 0$ and $S$ has minimum cardinality among the sets in $ \Omega_f$, we have $1 \leq \abs{S} \leq \abs{T}$.
\begin{itemize}
\item[Case]$2a$. $\abs{T} = 1$, which implies $\abs{S} = 1$. Suppose $S = \{s\}$ and $T = \{t\}$.  Consider a set $Q\subseteq [k]$ with $\abs{Q} = k - 2$. By the above assumptions, we have that if $s \in Q$ or $t \in Q$ then $Q \in \Omega_f$.
 This accounts for all but one sets $Q\subseteq [k]$ with $|Q|=k-2$; for the remaining set $Q=[k]\backslash \{s,t\}$, it must be the case that  $Q \not\in \Omega_f$, otherwise all sets $Q$ with $\abs{Q} = k - 2$ are in $\Omega_f$, which contradicts the fact that $f^* = f_{\mathsf{allone}}$. 
Now let's consider $\Omega_f$. 
The number of sets $W\in \Omega_f$ which contain both $s$ and $t$ is $2^{k-2}$.
Similarly, the number of sets $W\in \Omega_f$ which contain $s$ but not $t$ is $2^{k-2}$ 
 and the number of sets $W\in \Omega_f$ which contain $t$ but not $s$ is $2^{k-2}$.
 But the number of sets $W \in \Omega_f$ which contain neither~$s$ nor~$t$ is less than $2^{k-2}$.
 So the $k$-tuple hypergraph with the single hyperarc $(v_1,\hdots,v_k)$  
 induces a hard function on the two vertices~$v_s$ and~$v_t$
 and therefore $f$ simulates a hard function.

\item[Case]$2b$. $\abs{T} \geq 2$ and $S \cap T \neq \emptyset$. Since $S \setminus T \neq \emptyset$, we have $\abs{S} > \abs{S \cap T}$ and thus $S \cap T \not\in \Omega_f$ by the minimality of $S$. Let $r \in S \cap T$. Now we know that $S \in \Omega_f, T \in \Omega_f$ and $S \cup T \in \Omega_f$ by the assumptions above. But $S \cap T \not\in \Omega_f$ and $\chi_S \oplus \chi_T \oplus \chi_{S\cup T} = \chi_{S \cap T}$, so by Item~\ref{it:abc} of Lemma~\ref{lem:affine}, $f_{r\ra1}$ is not affine.

\item[Case]$2c$. $\abs{T} \geq 2$ and $S \cap T = \emptyset$. Since $T \setminus S \neq \emptyset$, let $r \in T \setminus S$. By the above assumptions, we have that $S \cup \{r\}$ and $S \cup T$ are in $\Omega_f$. Note that $\{r\}\notin \Omega_f$; otherwise, we would obtain a contradiction to the choice of the set $T$, since $T'=\{r\}$ satisfies $T'\in \Omega_f$, $S\backslash T'=S\neq \emptyset$ and $|T'|<|T|$. Now we know that $S \cup \{r\}, T, S \cup T \in \Omega_f$ and $\{r\}\not\in \Omega_f$. Note that since $S \cap T = \emptyset$, it holds that $\chi_{S\cup\{r\}}\oplus \chi_T \oplus \chi_{S\cup T} = \chi_{\{r\}}$, so by Item~\ref{it:abc} of Lemma~\ref{lem:affine} we have that $f_{r\ra1}$ is not affine.
\end{itemize}
\end{itemize}
This concludes the proof of Lemma~\ref{lem:allone}.
\end{proof}

Similarly, by switching the spins 0 and 1, we obtain the following lemma when $f^* = f_\mathsf{allzero}$.
\begin{lemma}\label{lem:allzero}
Let $k\geq 2$ and let
$f:\{0, 1\}^k \ra \{0, 1\}$ be a $k$-ary Boolean function.    Suppose that $f\neq f_\mathsf{allzero}$ and $f^*= f_\mathsf{allzero}$. Then at least one of the four following propositions is true:
\begin{enumerate} 
\item $f$ is semi-trivial;
\item there exists $t \in [k]$ such that $f_{t\ra0}$ is not affine;
\item $f$ supports  perfect equality;
\item $f$ simulates a hard function.
\end{enumerate}
\end{lemma}
\begin{proof} 
Suppose $f$ is a Boolean function such that $f^*=f_{\mathsf{allzero}}$ and $f \neq f_{\mathsf{allzero}}$. Let $g$ be the function defined by  $g(\xb) = f(\overline{\xb})$ for all $\xb\in\{0,1\}^k$.
Now it holds that $g^*=f_{\mathsf{allone}}$ and $g^*\neq f_\mathsf{allone}$. So $g$ satisfies one of the four propositions in Lemma~\ref{lem:allone}. We then have
\begin{enumerate}
\item If $g$ is semi-trivial, $f$ is semi-trivial.
\item If $g_{t\ra1}$ is not affine for some $t\in [k]$, $f_{t\ra 0}$ is not affine either.
\item If $g$ supports perfect equality, $f$ supports perfect equality too.
\item If $g$ simulates a hard function, $f$ simulates  the bitwise complement of the hard function, which is also hard.\qedhere
\end{enumerate}
\end{proof}

For every function $f$ such that $f^* = f_\mathsf{zero}$ and $f\neq f_\mathsf{zero}$, we still have a similar reduction lemma, but the proof is  more complicated.

\begin{lemma}\label{lem:zeroreduction}
Let $k\geq 3$ and let
$f:\{0, 1\}^k \ra \{0, 1\}$ be a $k$-ary Boolean function.  Suppose that $f^*= f_\mathsf{zero}$ and $f\neq f_\mathsf{zero}$. Let $S\in \Omega_f$. Then,  at least one of the four following propositions is true:
\begin{enumerate} 
\item $h=f_{\overline{S}\ra0}$ is semi-trivial;
\item \label{it:zeroaffine} there exists $T\subseteq [k]$ such that $f_{\overline{S}\ra0,T\ra 1}$ is not affine;
\item \label{it:zeroequality} $f$ supports perfect equality;
\item \label{it:zerohard} $f$ simulates a hard function.
\end{enumerate}
\end{lemma}

\begin{proof}
By Lemma~\ref{lem:zeropinning}, we have that either $f$ supports perfect equality or $f$ supports 
both  perfect pinning-to-$0$ and perfect pinning-to-$1$. We assume that the latter holds (otherwise we are done).

Let $h = f_{\overline{S}\ra 0}$. Since $f(\zeroes)=f^*(\zeroes)=0$ and $S\in \Omega_f$, we have that $h(\zeroes) = 0$ and $h(\ones) = 1$. 
Note that $h$ has arity $q:=\abs{S}$. We may assume that $q>1$; otherwise, $h$ is semi-trivial
(proposition~1 in the statement of the lemma). 
There are two cases to consider: $h^* \not\in \mathsf{EASY}(q)$ or $h^* \in \mathsf{EASY}(q)$. 
\begin{itemize} 
\item Case $1$. $h^* \not\in \mathsf{EASY}(q)$. 

\begin{description}
\item {\bf Case 1a. $q=2$.\quad}

In this case,  $h^*(0,0) =0$ and $h^*(0,1) =h^*(1,0)= h^*(1,1) = 1$, 
so $h^* = \mathsf{OR}$ which is a hard function. 
We have already assumed (in the first line of the proof) that $f$ supports perfect pinning-to-$0$.
Also, by definition, $h$ perfectly simulates itself.
By Item~3 of Lemma~\ref{lem:reduction}, $f$  perfectly simulates~$h$ as well, so $f$ simulates a hard function 
(proposition~4 in the statement of the lemma).

\item {\bf Case 1b. $q>2$.\quad}

 By Lemma~\ref{cor:corfsymm}, either $h$ simulates a hard function 
 or $h$ supports perfect equality (or both).
 If $h$ simulates a hard function then  by 
 Item~3 of Lemma~\ref{lem:reduction},   $f$ simulates the same hard function (proposition~4 in the statement of the lemma).
 On the other hand, if $h$ supports perfect equality then 
 by Item~1 of Lemma~\ref{lem:reduction}
 so does $f$ (proposition~3 in the statement of the lemma). 
 \end{description}

\item Case $2$. $h^* \in \mathsf{EASY}(q)$. Since $h(\zeroes) = 0$ and $h(\ones) = 1$, we have that $h^*$ is $f_\mathsf{odd}$ or $f_\mathsf{allone}$. 

\begin{description}
\item {\bf Case 2a. $h^* =  f_{\mathsf{odd}}$.\quad}  By Lemma~\ref{lem:parity}, $h$ supports perfect equality and thus $f$ supports perfect equality by 
Item~1 of Lemma~\ref{lem:reduction} (proposition~3 in the statement of the lemma).

\item {\bf Case 2b.  $h^* = f_{\mathsf{allone}}$.\quad} If $h = f_{\mathsf{allone}}$, $h$ is semi-trivial
(proposition~1 in the statement of the lemma). Otherwise, note that $q>1$, so by Lemma \ref{lem:allone},  
$h$ is semi-trivial (proposition~1 in the statement of the lemma),
or 
there exists $t\in [k]$ such that $h_{t\rightarrow 1}$ is not affine
or $h$ supports perfect equality or $h$ simulates a hard function.
If there exists $t\in [k]$ such that $h_{t\rightarrow 1}$ is not affine
then taking $T=\{t\}$,
$f$ satisfies proposition~2 in the statement of the lemma.
Finally, if $h$ supports perfect equality then so does~$f$ (like Case 2a)
and if $h$ simulates a hard function, then so does~$f$ (like Case 1b).\qedhere 
\end{description}
\end{itemize}
\end{proof}

\begin{lemma}\label{lem:zero}
Let $k\geq 2$ and let $f:\{0, 1\}^k \ra \{0, 1\}$ be a $k$-ary Boolean function. Suppose that $f\neq f_\mathsf{zero}$ and $f^*= f_\mathsf{zero}$. Then at least one of the four following propositions is true:
\begin{enumerate} 
\item $f$ is affine;
\item \label{it:zeroaffineb} there exist $S, T\subseteq [k]$ such that $f_{S\ra0,T\ra 1}$ is not affine;
\item \label{it:zeroequalityb} $f$ supports perfect equality;
\item \label{it:zerohardb} $f$ simulates a hard function.
\end{enumerate}
\end{lemma}
\begin{proof}
If $k = 2$, we have $f(0,0) = f(1,1) = 0$, so $\abs{\Omega_f}\leq 2$ and thus $f$ is affine (cf. Item~\ref{it:abc} of Lemma~\ref{lem:affine}) so it satisfies proposition~1 in the statement of the lemma. 

Now suppose $k \geq 3$.
By Lemma~\ref{lem:zeroreduction}, we can assume that for all $W \in \Omega_f$, $ f_{\overline{W}\ra 0}$ is a semi-trivial function (otherwise $f$ satisfies at least one of propositions~\ref{it:zeroaffine},~\ref{it:zeroequality} or~\ref{it:zerohard}). 

Choose $S \in \Omega_f$ such that $|S|$ is as large as possible. Let $h = f_{\overline{S}\ra 0}$. 
Since $h$ is semi-trivial (by taking $W=S$ above),  
we claim that
 there is a $T$ satisfying $\emptyset \subset T \subseteq S$
such that $\Omega_h = \{U \mid T \subseteq U \subseteq S\}$.  
(To see this, note that the definition of semi-trivial 
implies that there is a subset $T$ of $S$ such that
either $\Omega_h = \{U \mid U \subseteq T\}$ or $\Omega_h = \{U \mid T \subseteq U \subseteq S\}$.
The former  is impossible since $\emptyset \not\in \Omega_h$ since $h(\zeroes) = f(\zeroes)$
and $f(\zeroes)=0$ since $f^*= f_\mathsf{zero}$.
Also, in the latter case, $T$ is not empty because, once again, $\emptyset \not\not\in\Omega_h$.)

{\bf Case 1.} {\sl Suppose that $\forall X \in \Omega_f$, $T \subseteq X$:\quad}
Recall that $T$ is non-empty.
Also, for every $i\in T$, $\{i\} \cup \Omega_{f_{i \rightarrow 1}} = \Omega_f$ 
so either $f$ is affine (proposition~1 in the statement of the lemma)
or $f_{i\rightarrow 1}$ is not affine (proposition~2 in the statement of the lemma).

Now, if Case~1 does not hold
then there is an $X \in \Omega_f$ such that $T\setminus X$ is non-empty.
Since $\Omega_h = \{U \mid T \subseteq U \subseteq S\}$
we conclude that $X \notin \Omega_h$.
Since $h = f_{\overline{S}\ra 0}$
we conclude that $X\setminus S$ is non-empty.
Thus, the only other case to consider is as follows.

{\bf Case 2.} {\sl Suppose that there is an $X\in \Omega_f$ such that $T\setminus X$ and $X\setminus S$ are both non-empty:\quad}

Let $\Psi = \{X \in \Omega_f \mid \mbox{
$T\setminus X \neq \emptyset$ and $X\setminus S \neq \emptyset$
}\}$.
Let  
$a = \min \{|T\setminus X| : X\in \Psi\}$, and
$b = \min \{|X\setminus S| : \mbox{ $X\in \Psi$ and $|T\setminus X|=a$}\}$.
Choose $R\in \Psi$ with $|T\setminus R| =a$ and $|R \setminus S| = b$.

Now before proceeding, we use the sets $S$, $T$ and $R$ to partition~$k$.  
\begin{align*}
A &= \{ i\in [k] \mid i\in S, i \in T, i\notin R\},\\
B &= \{ i\in [k] \mid i\in S, i \in T, i\in R\},\\
C &= \{ i\in [k] \mid i\in S, i \notin T, i\notin R\},\\
D &= \{ i\in [k] \mid i\in S, i \notin T, i\in R\},\\
E &= \{ i\in [k] \mid i\notin S, i \notin T, i\notin R\},\\
F &= \{ i\in [k] \mid i\notin S, i \notin T, i\in R\}.
\end{align*}
It is clear from the definitions
that the sets $A$, $B$, $C$, $D$, $E$ and $F$ are disjoint.
Also, since $T\subseteq S$, they partition~$[k$].
 From the definitions, $A =T\setminus R$ 
and $F = R\setminus S$ so, by  the choice of $R$, $A$ and $F$
 are  non-empty.
 Let $g = f_{C\cup E \rightarrow 0, B\cup D \rightarrow 1}$.

By definition, every element of $\Omega_g$ is a subset of $A \cup F$. 
Also, for $Y \subseteq A \cup F$, ``$Y \in \Omega_g$'' means the same thing as
``$Y \cup B \cup D \in \Omega_f$''.
We establish   some facts before dividing the analysis into   sub-cases.

\begin{description}
\item [{\bf Fact 1: $A\in \Omega_g$.}]\quad
We have $\Omega_h = \{U \mid T \subseteq U \subseteq S\}$
and $T  = A \cup B$
so $A \cup B \cup D \in \Omega_h$.
Since $A \cup B \cup D \subseteq S$,
this means $A \cup B \cup D \in \Omega_f$.
Equivalently, $A\in \Omega_g$.

\item [{\bf Fact 2: $F\in \Omega_g$.}]\quad
From the definition of $R$, $R\in \Omega_f$.
Also, $R= B \cup D \cup F$ 
so $F \cup B \cup D \in \Omega_f$.
Equivalently, $F \in \Omega_g$.

\item [{\bf Fact 3:  If $Y\in \Omega_g$ 
then  either
$Y \cap A \in \{\emptyset,A\}$ or
$Y \cap F  =\emptyset$ (or both).}]\quad
Suppose for contradiction that 
$\emptyset \subset Y \cap A \subset A$
and 
$  Y \cap F $ is non-empty.
Note that $R = B \cup D \cup F$.
Let $R' = B \cup D \cup  Y$.
Note that $T\setminus R = A$
and $T \setminus R' = A \setminus Y \subset A$
so $|T \setminus R'| < |T \setminus R|$.
We will show a contradiction to the choice of~$R$ by showing that $R'\in \Psi$.
First, since $Y\in \Omega_g$, $R' \in \Omega_f$.
Also, $T\setminus R' = A \setminus Y$ is non-empty
  and $R'\setminus S = Y \cap F$  is non-empty.

\item [{\bf Fact 4:  If $Y\in \Omega_g$  
and $Y \cap A = \emptyset$ then $Y \in \{\emptyset,F\}$.}]\quad
Suppose for contradiction that 
$\emptyset \subset Y \subset F$.   As in the proof of Fact~3, let
$R' = B \cup D \cup  Y$. 
Note that $T\setminus R = T\setminus R' = A$.
Also, $R\setminus S = F$ and
$R'\setminus S = Y$
so $|R\setminus S| > |R'\setminus S|$. 
Once again, we will show a contradiction to  the choice of~$R$ by showing that $R'\in \Psi$.
As in the proof of Fact~3, since $Y\in \Omega_g$, $R' \in \Omega_f$.
Also, $T\setminus R'$ is non-empty since $T\setminus R$ is.  
Finally,
$R'\setminus S =  Y$, which  is non-empty.

\item [{\bf Fact 5:  If $Y\in \Omega_g$ and $Y  \cap F = \emptyset$ then $Y=A$.}]\quad
Since $Y \in \Omega_g$, we have
$Y \cup B \cup D \in \Omega_f$.
But since
 $Y \subseteq A$, we have $Y \cup B \cup D \subseteq S$,
 so $Y \cup B \cup D \in \Omega_h$.
Since
$\Omega_h = \{U \mid T \subseteq U \subseteq S\}$
we have $T \subseteq Y \cup B \cup D$ so 
$A \subseteq Y$.
\end{description}

Given Facts~1--5, we have only the following sub-cases.
\begin{description}
\item [{\bf  Case 2a: $\Omega_g = \{A,F\}$.}]\quad
In this case, we will show that $f$ supports perfect equality
so it satisfies proposition~3 in the statement of the lemma.
Using Lemma~\ref{lem:zeropinning}, we conclude that either $f$ supports perfect equality 
(in which case we are finished)
or $f$ supports perfect pinning-to-$0$ and also perfect pinning-to-$1$, which we now assume.
Let $H_0$ be a $k$-tuple hypergraph, with a vertex $u_0$ such that $\mu_{f;H_0}(\sigma_{u_0}=0)=1$.
Let $H_1$ be a  $k$-tuple hypergraph, with a vertex $u_1$ such that $\mu_{f;H_1}(\sigma_{u_1}=0)=1$.
We have already noted that $A$ is non-empty. Suppose, without loss of generality, that $1\in A$
(otherwise, we simply re-order the arguments of $[k]$).
Now let $H'$ be the $k$-tuple hypergraph with vertices $v_0,v_1,\ldots,v_k$
and hyperarcs $(v_0,v_2,\ldots,v_k)$ and $(v_1,v_2,\ldots,v_k)$.
Construct~$H$ from $H'$ by doing the following:
\begin{itemize}
\item For every $i\in C \cup E$, take a new copy of $H_0$ and identify vertex $u_0$ with $v_i$.
\item For every $i\in B \cup D$, take a new copy of $H_1$ and identify vertex $u_1$ with $v_i$.
\end{itemize}
Now since $\Omega_g= \{A,F\}$, $\mu_{f;H}(\sigma(v_0)=\sigma(v_1)=0) = 
\mu_{f;H}(\sigma(v_0)=\sigma(v_1)=1)=1/2$.
Thus, $f$ supports perfect equality, so we have finished  Case 2a.

\item [{\bf  Case 2b:  $\exists Y \in \Omega_g$ such that 
$Y\cap A = A$ and $Y\cap F$ is non-empty.}]\quad 
We will show that $f$ satisfies proposition~2 in the statement of the lemma.
Specifically, consider some $t\in A$.
We will show that $f_{t\rightarrow 1}$ is not affine. 

Let $Y' = Y \cap F$ so that
$Y = A \cup Y'$. Let 
\[S_1 := B \cup D \cup Y = B \cup D \cup A \cup Y',\quad S_2 := A \cup B=T, \quad S_3 := A \cup B \cup C.\]
We claim that $S_1\backslash\{t\}, S_2\backslash\{t\}, S_3\backslash\{t\}\in \Omega_{f_{t\rightarrow 1}}$. Since $t\in S_1,S_2,S_3$ (from $t\in A$), the claim will follow by showing that  $S_1,S_2,S_3\in \Omega_f$. Indeed, since $Y\in \Omega_g$, we have that $S_1 \in \Omega_f$. Also, since $S_2=T$, we have that $S_2 \in \Omega_h$ so $S_2 \in \Omega_f$. Finally, since $T=A\cup B\subseteq  S_3 \subseteq A\cup B\cup C\cup D=S$, we have that $S_3 \in \Omega_h$ so $S_3 \in \Omega_f$.

Let $S':= A \cup B \cup C \cup D \cup Y'=S\cup Y'$ and note that $\chi_{S'}=\chi_{S_1}\oplus \chi_{S_2}\oplus \chi_{S_3}$ (see Section~\ref{sec:notation} for the relevant notation) by the disjointness of $A,B,C,D,E,F$.  Since $Y'$ is non-empty by assumption, we obtain that $S'$ is not in $\Omega_f$
by maximality of~$S$. Note that $t\in S'$,  so we have that $S'\backslash\{t\}\notin \Omega_{f_{t\rightarrow 1}}$. 

To sum up, we have shown that
\[S_1\backslash\{t\}, S_2\backslash\{t\}, S_3\backslash\{t\}\in \Omega_{f_{t\rightarrow 1}} \mbox{ but } S'\backslash\{t\}\notin \Omega_{f_{t\rightarrow 1}}\]
Since $\chi_{S_1\backslash\{t\}}\oplus \chi_{S_2\backslash\{t\}}\oplus \chi_{S_3\backslash\{t\}}=\chi_{S'\backslash\{t\}}$, by Item~\ref{it:abc} of Lemma~\ref{lem:affine}, $f_{t\rightarrow 1}$ is not affine. 
\end{description}
This concludes the proof of Lemma~\ref{lem:zero}.  
\end{proof}

Now we can prove Theorem \ref{thm:asym-main}, which we restate here for convenience.
\begin{thmasymmain}
\statethmasymmain{} 
\end{thmasymmain}
\begin{proof} 
We prove this Theorem by induction on the arity of $f$.

\begin{itemize} 
\item $k = 2$. So $R_f \subseteq \{00, 01, 10, 11\}$. If $f$ is not affine, then $\abs{R_f} = 3$.

If $00 \not\in R_f$ or $11 \not\in R_f$, let $G$ be a graph with two vertices $u$ and $v$ and an edge $(u, v)$. Then either $\mu_{f;G}(\sigma_u = 1, \sigma_v = 1)=\mu_{f;G}(\sigma_u = 0, \sigma_v = 1) = \mu_{f;G}(\sigma_u= 1, \sigma_v = 0) = \frac{1}{3}$ or  $\mu_{f;G}(\sigma_u = 0, \sigma_v = 0)=\mu_{f;G}(\sigma_u = 0, \sigma_v = 1) = \mu_{f;G}(\sigma_u = 1, \sigma_v = 0) = \frac{1}{3}$. So $f$ simulates a hard function.

If $01 \not\in R_f$ or $10 \not\in R_f$, $f^*$ will be $f_{\mathsf{EQ}}$ and thus $f$ supports  perfect equality by Lemma~\ref{lem:equalizer}.

\item $k \geq 3$. Suppose that for all $2 \leq k' < k$, all $k'$-ary functions  $f'$ satisfy at least one of the three propositions in the statement. We now prove that an arbitrary $f:\{0,1\}^k\rightarrow \{0,1\}$ satisfies at least one of the propositions as well.
If $f$ is affine, then it satisfies proposition~1 in the statement of the lemma,
so we assume that $f$ is not affine, so $f \not\in \mathsf{EASY}(k)$.
 We have the following case analysis.

\begin{enumerate} 
\item[Case]$1$. $f^* \not\in \mathsf{EASY}(k)$. By Lemma~\ref{cor:corfsymm}, $f^*$ either simulates a hard function in which case $f$  simulates the same hard function as well or $f^*$ supports perfect equality in which case $f$ supports perfect equality as well.

\item[Case]$2$. $f^* \in \mathsf{EASY}(k)$. There are $6$ sub-cases to consider:
\begin{itemize} 
\item[Case]$2a$. $f^* = f_\mathsf{EQ}$. By Lemma~\ref{lem:equalizer}, $f$ supports  perfect equality.

\item[Case]$2b$. $f^* = f_\mathsf{odd}$ or $f^* = f_\mathsf{even}$. By Lemma~\ref{lem:parity}, $f$ supports perfect equality.

\item[Case]$2c$. $f^* = f_\mathsf{allone}$. By Lemma~\ref{lem:allonepinning}, $f$ supports perfect pinning-to-$1$. By Lemma~\ref{lem:allone}, $f$ is semi-trivial (and thus $f$ is affine), or $f$ supports perfect equality or simulates a hard function, or there exists $t \in [k]$ such that $f_{t\ra 1}$ is not affine. If $f_{t \ra 1}$ is not affine for some $t\in [k]$, $f_{t\ra 1}$ must support perfect equality or simulate a hard function by the induction hypothesis. 
So $f$ supports or simulates the same function by   Lemma~\ref{lem:reduction}. 
								
\item[Case]$2d$. $f^* = f_\mathsf{allzero}$. The proof for this case is completely analogous to the case $2c$ by switching the spins 0 and 1 (cf. Lemma~\ref{lem:allzero}).

\item[Case]$2e$. $f^* = f_\mathsf{one}$. $f^*= f_\mathsf{one}$ means $f(\xb)=1$ for all $\xb \in \{0, 1\}^k$, so $f$ is affine.

\item[Case]$2f$. $f^* = f_\mathsf{zero}$. By Lemma~\ref{lem:zeropinning}, we have that either $f$ supports perfect equality or $f$ supports   both perfect pinning-to-$0$ and perfect pinning-to-$1$. We assume that the latter is the case (otherwise we are done). By Lemma~\ref{lem:zero}, $f$ is affine, or $f$ supports perfect equality or simulates a hard function, or there exists some 
$S,T \subseteq [k]$ such that $f_{S\rightarrow 0, T \rightarrow 1}$ is not affine.
The only case where we aren't immediately finished is the final one.
In this case,  the arity of $f_{S\rightarrow 0, T \rightarrow 1}$ must be at least~$2$ since every 
unary function is affine.
Thus, since  $f_{S\rightarrow 0, T \rightarrow 1}$ is not affine, 
  it must support perfect equality or simulate a hard function $g$ by the induction hypothesis.
 Then, $f$ either supports perfect equality 
 or simulates the hard function~$g$ by Lemma~\ref{lem:reduction}. 
\end{itemize}
\end{enumerate}
\end{itemize}
This concludes the case analysis and the proof of Theorem~\ref{thm:asym-main}.
\end{proof}

\section{The case where $f$ supports perfect equality}
\label{sec:equalitysec}

In this section, we assume that $f$ is not affine but that it supports perfect equality. In this case, due to the presence of perfect equality, we will be able to employ results and techniques from \cite{trichotomy} to show Theorem~\ref{thm:perfect}.

\subsection{Constraint Satisfaction Problems and Implementations}\label{sec:aaa}

In the introduction to this paper, we illustrated how Boolean relations can
implement more complicated interactions
by considering the ``not-all-equal'' relation of arity~$3$ and using it to ``implement'' ferromagnetic Ising
interactions. At this point, it is useful to make the notion of ``implement''
more precise. There are various notions in the literature of \emph{implementations}.
We use (a generalisation of) the notion from~\cite{trichotomy}, which is  
essentially the ``faithful, perfect'' variant of ``implementation'' from~\cite{CKSb}.

 \begin{definition} \label{def:implementation}
Let $\Gamma$ be a Boolean constraint language. The language $\Gamma$ \emph{implements} a $t$-ary function $g:\{0, 1\}^t \ra \Rplus$, if for some $t'\geq t$ there is a CSP instance $I$ with variables $x_1,\hdots,x_{t'}$ and constraint language $\Gamma$ such that for every tuple $(s_1,\hdots,s_t)\in \{0,1\}^t$, there are precisely $g(s_1,s_2,\hdots,s_t)$ satisfying assignments $\sigma$ of $I$ with $\sigma(x_1)=s_1,\hdots, \sigma(x_t)=s_t$.\footnote{See also the relevant equation~\eqref{eq:rf3454frg5}.}
\end{definition}

When $f$ supports perfect equality, we
will use the following ``transitivity'' lemma, which will allow us to use some known implementations.

\begin{lemma}\label{lem:transitivity}
Let $f:\{0,1\}^k\rightarrow \{0,1\}$ be a Boolean function which supports perfect equality. Let $\Gamma$ be a 
Boolean constraint language and let $g:\{0, 1\}^t \ra \Rplus$ be a $t$-ary function 
such that $g$ is not $\zerof^{(t)}$ and $\Gamma$ implements $g$. Then, if $f$ perfectly simulates  $\Gamma$, $f$ also perfectly simulates the function $g$. 
\end{lemma}

\begin{proof} 
Since $\Gamma$ implements $g$, there exists some $t'\geq t$ and a CSP instance $I$ with variables $X:=\{x_1,\hdots,x_{t'}\}$ and constraints in $\Gamma$ such that for every tuple $(s_1,\hdots,s_t)\in \{0,1\}^t$, there are precisely $g(s_1,s_2,\hdots,s_t)$ satisfying assignments $\sigma$ of $I$ with $\sigma(x_1)=s_1,\hdots, \sigma(x_t)=s_t$. 
 Since $g\neq \zerof^{(t)}$, we conclude that,
   for all $s_1,\hdots,s_t\in \{0,1\}$, 
\begin{equation}\label{eq:rf3454frg5}
 \mu_{I}(\sigma(x_1)=s_1,\hdots, \sigma(x_t)=s_t)=\frac{g(s_1, s_2, \ldots, s_t)}{\sum\limits_{(s_1', s_2', \ldots, s_t') \in \{0, 1\}^t}g(s_1', s_2', \ldots, s_t')}.
 \end{equation}   

Since $f$ supports perfect equality, there exists a  $k$-tuple hypergraph $H_{\eq}=(V_\eq,\mathcal{F}_\eq)$ and vertices $y,z\in V_\eq$ such that
\begin{equation}\label{eq:Heqyz}
\mu_{f;H_{\eq}}(\sigma(y)=\sigma(z)=0)=\mu_{f;H_{\eq}}(\sigma(y)=\sigma(z)=1)=1/2.
\end{equation}
        
We will use the CSP instance $I$ and the hypergraph $H_{\eq}$ to construct a  $k$-tuple hypergraph $H=(V,\mathcal{F})$ with vertices $v_1,\hdots,v_{t'}$ in $V$ satisfying
\begin{equation}\label{eq:goaledc}
\mu_{f;H}(\sigma(v_1)=s_1,\hdots,\sigma(v_{t'})=s_{t'})=\mu_{I}(\sigma(x_1)=s_1,\hdots, \sigma(x_{t'})=s_{t'})
\end{equation}
for all $s_1,\hdots, s_{t'}\in\{0,1\}$. From this, the lemma follows since we can sum over the values of $s_{t+1},\hdots,s_{t'}\in \{0,1\}$ to obtain that 
\[\mu_{f;H}(\sigma(v_1)=s_1,\hdots,\sigma(v_{t})=s_{t})=\mu_{I}(\sigma(x_1)=s_1,\hdots, \sigma(x_{t})=s_{t})\]
for all $s_1,\hdots, s_{t}\in\{0,1\}$, which in conjuction with \eqref{eq:rf3454frg5} yields that $f$ perfectly simulates $g$.

To formally construct the  $k$-tuple hypergraph $H$, we will need some notation. 
Suppose that $I$ has $m$ constraints
and for $j\in[m]$ write the $j$'th constraint
as $f_j(x_{j,1},\ldots,x_{j,w(j)})$, 
where $w(j)$ is the arity of $f_j\in \Gamma$
and, for all $i\in[w(j)]$, $x_{j,i} \in \{x_1,\ldots,x_{t'}\}$.
Since $f$ perfectly simulates $\Gamma$ and every~$f_j$ is  in~$\Gamma$, 
for every constraint  
$C_j = f_j(x_{j,1},\ldots,x_{j,w(j)})$,  
there is a  $k$-tuple hypergraph $H_j=(V_j,\mathcal{F}_j)$ and vertices $v_{j,1},\hdots,v_{j,w(j)}$ of $H_j$ such that for all $s_1,\hdots,s_{w(j)}\in \{0,1\}$, it holds that
\begin{equation}\label{eq:fsupportsfj}
\mu_{f;H_j}(\sigma(v_{j,1})=s_1,\hdots,\sigma(v_{j,w(j)})=s_{w(j)})=
\frac{f_j(s_1,\hdots,s_{w(j)})}
{|R_{f_j}|}.
\end{equation}   
Note that the expression $|R_{f_j}|$ in the denominator in~\eqref{eq:fsupportsfj} is
not zero  because the constraint~$C_j$ has a satisfying assignment, since $I$ does
(which follows from the fact that $g\neq \zerof^{(t)}$ and from the definition of~$I$).
        
Consider now the  $k$-tuple hypergraph $H'=(V',\mathcal{F}')$ which is simply the disjoint union of   $H_1,\hdots, H_m$ (i.e., $V'=\cup^m_{j=1}V_j$ and $\mathcal{F}'=\cup^m_{j=1}\mathcal{F}_j$). Note that, for every subset $S\subseteq V'$ and every assignment $\tau:S\rightarrow \{0,1\}$, it holds that 
\begin{equation}\label{eq:disjointness}
\mu_{f;H'}(\sigma_S=\tau)=\prod^m_{j=1}\mu_{f;H_j}(\sigma_{S\cap V_j}=\tau_{S\cap V_j}).
\end{equation}
To complete the construction of the desired~$H$, we need some further notation. For a variable $x_i\in \{x_1,\hdots,x_{t'}\}$ of the  CSP instance $I$, let $U_i\subseteq V'$ denote the subset of vertices of~$H'$ which correspond to  occurrences of the variable $x_i$ in the CSP instance $I$. 
More precisely, assume that the variable $x_i$ has $d$ occurrences in $I$ for some integer $d\geq 1$, and let $C_{j_1},\hdots,C_{j_d}$ be the constraints  in which $x_i$ appears (note that the indices $j_1,\hdots,j_d$ are not necessarily distinct). Further, let $t_1,\hdots,t_d$ denote the indices of the positions where $x_i$ appears in $C_{j_1},\hdots,C_{j_d}$ respectively. Then $U_i:=\{v_{j_1,t_1},\hdots,v_{j_d,t_d}\}$ is precisely the subset of vertices of~$H'$ 
which correspond to   occurrences of the variable $x_i$ in the CSP instance $I$.  Let $U:=\cup_{i\in[t']} U_i$ (note that in general $U\neq V'$ since $H'$ may contain vertices that do not correspond to  occurrences  of variables of $I$).
        
We are now ready to complete the construction of the desired  $k$-tuple hypergraph $H=(V,\mathcal{F})$. 
Start by setting $H$ equal to   $H'$.
Then, for
 each $i\in [t']$ and each pair of vertices $u,u'\in U_i$, add to $H$ a distinct copy of the 
  $k$-tuple hypergraph $H_\eq$,
 identifying the vertices $y$ and $z$ of $H_\eq$ with the vertices $u$ and $u'$. 
Having defined~$H$,
we next choose the  specified vertices $v_1,\hdots,v_{t'}$.
In fact, it suffices, for each $i\in [t]$, to let $v_i$ be an arbitrary vertex in $U_i$.

It remains to prove that \eqref{eq:goaledc} holds.
We call an assignment $\tau:U\rightarrow\{0,1\}$ \emph{relevant} if for every $i\in[t']$ there exists $s_i\in\{0,1\}$ such that for every vertex $v\in U_i$, it holds that $\tau(v)=s_i$. For relevant assignments $\tau$, we will refer to the tuple $(s_1,\hdots,s_{t'})$ as the CSP assignment corresponding to $\tau$.  For non-relevant $\tau$, the copies of $H_{\eq}$ on top of the sets $U_1,\hdots,U_{t'}$ ensure that  $\mu_{f;H}(\sigma_{U}=\tau)=0$. For all relevant $\tau:U\rightarrow \{0,1\}$, we have from \eqref{eq:Heqyz} that 
\[\mu_{f;H}(\sigma_{U}=\tau)=\mu_{f;H'}(\sigma_{U}=\tau)\] 
and, hence, using \eqref{eq:disjointness}, we have that 
\begin{equation}\label{eq:5r4r54r5}
\mu_{f;H}(\sigma_{U}=\tau)=\prod^m_{j=1}\mu_{f;H_j}(\sigma_{U\cap V_j}=\tau_{U\cap V_j}).
\end{equation}
Note that for every $j\in [m]$ we have $U\cap V_j=\{v_{j,1},\hdots,v_{j,w(j)}\}$ and, hence, \eqref{eq:fsupportsfj} gives
\begin{equation}\label{eq:epspselect}
\mu_{f;H_j}(\sigma_{U\cap V_j}=\tau_{U\cap V_j})=\frac{
f_j( \tau(v_{j,1}),\cdots,\tau(v_{j,w(j)}))}
 {|R_{f_j}|}.
\end{equation}
It follows from \eqref{eq:5r4r54r5} and \eqref{eq:epspselect} that
\begin{equation}\label{eq:epspselect2b}
\mu_{f;H}(\sigma_{U}=\tau)\propto\prod^m_{j=1}
f_j( \tau(v_{j,1}),\cdots,\tau(v_{j,w(j)}))
  \mbox{ for all relevant $\tau$}.
\end{equation}
For a relevant $\tau:U\rightarrow \{0,1\}$, let $(s_1,\hdots,s_{t'})$ be the CSP assignment corresponding to $\tau$. Then, the product in the r.h.s. of \eqref{eq:epspselect2b} is 1 iff  $(s_1,\hdots,s_{t'})$ encodes a satisfying assignment of the CSP instance $I$. Since the relevant $\tau:U\rightarrow \{0,1\}$ and assignments to the CSP instance $I$ are in 1-1 correspondence, we obtain \eqref{eq:goaledc}, as wanted.         
This concludes the proof of Lemma~\ref{lem:transitivity}.
\end{proof}

\begin{lemma}\label{lem:fgpinning}
Let $f:\{0, 1\}^k \ra \{0,1\}$ and $g:\{0, 1\}^t \ra \{0,1\}$ be Boolean functions such that $f$ simulates~$g$, and $g\neq 
\zerof^{(t)}$. 
Suppose that  $g$ supports pinning-to-$s$ for some $s\in \{0,1\}$. Then $f$ supports pinning-to-$s$ as well. 
\end{lemma}
\begin{proof}
Without loss of generality, we assume that $s=0$. Suppose that the function $g$ supports pinning-to-$0$. Our goal is to show that $f$ supports pinning-to-$0$ as well. 

First, let $Z_g :=\sum_{(s_1, s_2, \ldots, s_t) \in \{0, 1\}^t}g(s_1, s_2, \ldots, s_t)$.
Since $g\neq \zerof^{(t)}$, $Z_g>0$.
Since $g$ supports pinning-to-$0$, by Definition~\ref{def:boundedsupport1}, there exists a $t$-tuple hypergraph $H_0=(V_0,\mathcal{F}_0)$ and a vertex $v_0\in V_0$  such that
\begin{equation}\label{eq:gH09}
\mu_{g;H_0}(\sigma_{v_0}=0)\geq 9/10.
\end{equation}
(The choice of the constant $9/10$ is arbitrary, any constant greater than~$1/2$ would work.
Also, $Z_{g;H_0}>0$.)
For all $\eta: V_0 \rightarrow \{0,1\}$
define 
$A_\eta:=\frac{w_{g;H_0}(\eta)}{(Z_g)^{|\mathcal{F}_0|}}$
and define $M:= \sum_{\eta:V_0\rightarrow\{0,1\}} A_\eta = \frac{Z_{g;H_0}}{(Z_g)^{|\mathcal{F}_0|}}$.
Since $Z_{g;H_0}$ and $Z_g$ are positive, $M>0$. 
Also, \begin{equation}\label{eq:eFo3first}
\mu_{g;H_0}(\eta)=\frac{A_\eta}{M}.
\end{equation}
 
Now let $\epsilon:=\min\{M/8,1/8\}$, $\epsilon_1:=\epsilon/(|\mathcal{F}_0|\, 2^{2|V_0|})$ and $\epsilon_2:=\epsilon/( 2^{|V_0|})$. Since $f$ simulates the function $g$, by Definition~\ref{def:simulate} and Lemma~\ref{lem:generalpinfull},   there exists a $k$-tuple hypergraph $H_g=(V_g,\mathcal{F}_g)$ and $t$ vertices $v_1, v_2, \ldots, v_t$ of~$H_g$ such that, for all $(s_1, s_2, \ldots, s_t) \in \{0, 1\}^t$,
\begin{equation}\label{eq:Hg}
\Big|\mu_{f;H_g}(\sigma(v_1) = s_1, \sigma(v_2) = s_2,\ldots,\sigma(v_t) = s_t) -\frac{g(s_1, s_2, \ldots, s_t)}{Z_g}\Big|\leq \epsilon_1.
\end{equation}

Construct the $k$-tuple hypergraph $H=(V,\mathcal{F})$ as follows. For every 
hyperarc $e$ of $H_0$, say $e=(u_1,\hdots,u_t)\in \mathcal{F}_0$, take a distinct copy of $H_g$, which we will denote by $H^{(e)}_g$, and identify the vertices $u_1,\hdots,u_t\in V_0$ with the vertices $v_1,\hdots,v_t$ of $H_g$. Note that $H$ is a union of copies of $H_g$ which intersect only at vertices in $V_0$. 
Now for all $\eta: V_0 \rightarrow \{0,1\}$
define $A'_\eta:=\prod_{e\in \mathcal{F}_0}\mu_{f;H^{(e)}_g}(\sigma_{e}=\eta_e)$
and $M' := \sum_{\eta:V_0\rightarrow\{0,1\}} A'_{\eta}$.
 Then
\begin{equation}\label{eq:eFo0}
\mu_{f;H}(\sigma_{V_0}=\eta)= \frac{ A'_\eta}{ M'}.
\end{equation}
By \eqref{eq:Hg}, for every $e=(u_1,\hdots,u_t)\in \mathcal{F}_0$, it holds that  
\begin{equation}\label{eq:xsq23e}
\Big|\mu_{f;H^{(e)}_g}(\sigma_{e}=\eta_e)-\frac{g\big(\eta(u_1),\hdots,\eta(u_t)\big)}{Z_g}\Big|\leq \epsilon_1.
\end{equation}
Recall that for real numbers $a_1,\hdots,a_n\in [0,1]$ and $b_1,\hdots,b_n\in[0,1]$, it holds that $|\prod^n_{i=1}a_i-\prod^n_{i=1}b_i|\leq \sum^n_{i=1}|a_i-b_i|$. Thus, using \eqref{eq:xsq23e}, we obtain that, for every $\eta:V_0\rightarrow\{0,1\}$, it holds that 
\begin{equation}\label{eq:eFo1}
|A'_\eta - A_\eta| = 
\Big|\prod_{e\in \mathcal{F}_0}\mu_{f;H^{(e)}_g}(\sigma_{e}=\eta_e)-\frac{w_{g;H_0}(\eta)}{(Z_g)^{|\mathcal{F}_0|}}\Big|\leq \epsilon_1 |\mathcal{F}_0|.
\end{equation}
Summing this over all $\eta:V_0\rightarrow\{0,1\}$, we obtain that
\begin{equation}\label{eq:eFo2}
|M' - M| = \Big|\sum_{\eta:V_0\rightarrow\{0,1\}} A'_\eta-
\sum_{\eta: V_0 \rightarrow\{0,1\}} A_\eta \Big|\leq \epsilon_1 |\mathcal{F}_0| 2^{|V_0|}=\epsilon_2.
\end{equation}
 Note that the expression $\epsilon_1 |\mathcal{F}_0|$ in 
 \eqref{eq:eFo1} is at most~$\epsilon_2$. Also, 
for all $\eta: V_0 \rightarrow \{0,1\}$,
$A_\eta \leq M$ and $A'_\eta \leq M'$.
The bounds in \eqref{eq:eFo1} and \eqref{eq:eFo2} yield that $A_\eta-\epsilon_2\leq A'_\eta\leq A_\eta+\epsilon_2$ and $M-\epsilon_2\leq M'\leq M+\epsilon_2$.  
 By the choice of $\epsilon$, we have $M-\epsilon>M/2$ and hence $M-\epsilon_2>M/2$ as well. Further, we have the bound 
  \begin{equation}\label{eq:eFo4}
\Big|\frac{A_\eta}{M}-\frac{A'_\eta}{M'}\Big|\leq\max\Big\{\frac{A_\eta}{M}-\frac{A_\eta-\epsilon_2}{M+\epsilon_2},\frac{A_\eta+\epsilon_2}{M-\epsilon_2}-\frac{A_\eta}{M}\Big\}\leq \frac{\epsilon_2(A_\eta+M)}{M(M-\epsilon_2)}\leq \frac{2\epsilon_2}{M-\epsilon_2}\leq \frac{4\epsilon_2}{M}\leq \frac{1}{2^{|V_0|+1}}.
\end{equation}
 From \eqref{eq:eFo3first} and \eqref{eq:eFo0} and \eqref{eq:eFo4}, we thus obtain that  for every $\eta:V_0\rightarrow\{0,1\}$, it holds that 
\[|\mu_{f;H}(\sigma_{V_0}=\eta)-\mu_{g;H_0}(\eta)|\leq 1/2^{|V_0|+1}.\]
Summing this over the $2^{|V_0|-1}$ possible values of $\eta_{V_0\backslash \{v_0\}}$, we obtain that for $s\in \{0,1\}$ it holds that
\[|\mu_{f;H}(\sigma_{v_0}=s)-\mu_{g;H_0}(\sigma_{v_0}=s)|\leq 1/4.\]
Combining this with \eqref{eq:gH09}, we obtain that 
\[\mu_{f;H}(\sigma_{v_0}=0)>1/2>\mu_{f;H}(\sigma_{v_0}=1).\]
Thus, by Lemma~\ref{lem:gadgets}, we obtain that $f$ supports pinning-to-0. This concludes the proof of Lemma~\ref{lem:fgpinning}.
\end{proof}

The following lemma is similar to Lemma~\ref{lem:transitivity}
except that, instead of assuming that $f$ perfectly simulates~$\Gamma$, we only assume that
$f$ simulates~$\Gamma$ so instead of concluding that $f$ perfectly simulates~$g$,
we only conclude that $f$ simulates~$g$.

\begin{lemma}\label{lem:transitivityb}
Let $f:\{0,1\}^k\rightarrow \{0,1\}$ be a Boolean function which supports 
equality. Let $\Gamma$ be a 
Boolean constraint language and let $g:\{0, 1\}^t \ra \Rplus$ be a $t$-ary function 
such that $g$ is not $\zerof^{(t)}$ and
$\Gamma$ implements $g$. Then, if $f$   simulates  $\Gamma$, $f$ also   simulates the function $g$. 
\end{lemma}

\begin{proof}
The proof is similar to the proof of Lemma~\ref{lem:transitivity},  but the imperfect nature
of the simulation adds technical details.
Since $\Gamma$ implements~$g$ we can follow
the proof of Lemma~\ref{lem:transitivity} to define the CSP instance~$I$ 
with variables $\{x_1,\ldots,x_{t'}\}$ 
and constraints in $\Gamma$ 
satisfying~\eqref{eq:rf3454frg5}.

We will use the CSP instance $I$ and the fact that $f$ supports equality to construct a $k$-tuple  hypergraph $H=(V,\mathcal{F})$  with an admissible set $\mathcal{V}^*$  
for $H$ with respect to~$f$
 and vertices $v_1,\hdots,v_{t'}$ in $V$ satisfying
\begin{equation}\label{eq:goaledc12}
\mu_{f;H}^{\mathrm{cond}(\mathcal{V}^*)}(\sigma(v_1)=s_1,\hdots,\sigma(v_{t'})=s_{t'})=\mu_{I}(\sigma(x_1)=s_1,\hdots, \sigma(x_{t'})=s_{t'})
\end{equation}
for all $s_1,\hdots, s_{t'}\in\{0,1\}$. 
From this, the lemma follows since we can sum over the values of $s_{t+1},\hdots,s_{t'}\in \{0,1\}$ to obtain that 
\[\mu_{f;H}^{\mathrm{cond}(\mathcal{V}^*)}(\sigma(v_1)=s_1,\hdots,\sigma(v_{t})=s_{t})=\mu_{I}(\sigma(x_1)=s_1,\hdots, \sigma(x_{t})=s_{t})\]
for all $s_1,\hdots, s_{t}\in\{0,1\}$, which in conjuction with \eqref{eq:rf3454frg5} yields that $f$ simulates $g$.

To formally construct the $k$-tuple hypergraph $H$,  we will need some notation. 
As in the proof of Lemma~\ref{lem:transitivity}, suppose that $I$ has $m$ constraints
and for $j\in[m]$ write the $j$'th constraint
as $f_j(x_{j,1},\ldots,x_{j,w(j)})$, 
where $w(j)$ is the arity of $f_j\in \Gamma$
and, for all $i\in[w(j)]$, $x_{j,i} \in \{x_1,\ldots,x_{t'}\}$.
Since $f$ simulates $\Gamma$ and every~$f_j$ is  in~$\Gamma$, 
for every constraint  
$C_j = f_j(x_{j,1},\ldots,x_{j,w(j)})$,  
there is a $k$-tuple hypergraph $H_j=(V_j,\mathcal{F}_j)$, an admissible collection $\mathcal{V}^j=(V^j_{\pinzero},V^j_{\pinone},\mathcal{V}^j_{\eq})$ for $H_j$ with respect to $f_j$ and vertices $v_{j,1},\hdots,v_{j,w(j)}$ of $H_j$ such that for all $s_1,\hdots,s_{w(j)}\in \{0,1\}$, it holds that
\begin{equation}\label{eq:fsupportsfj12}
\mu_{f;H_j}^{\mathrm{cond}(\mathcal{V}^j)}(\sigma(v_{j,1})=s_1,\hdots,\sigma(v_{j,w(j)})=s_{w(j)})=
\frac{f_j(s_1,\hdots,s_{w(j)})}
{|R_{f_j}|}.
\end{equation}   
        
Consider now the $k$-tuple hypergraph $H=(V,\mathcal{F})$ which is simply the disjoint union of $H_1,\hdots, H_m$ (i.e., $V=\cup^m_{j=1}V_j$ and $\mathcal{F}=\cup^m_{j=1}\mathcal{F}_j$). Further,  let $\mathcal{V}=(V_{\pinzero},V_{\pinone},\mathcal{V}_{\eq})$, where $V_{\pinzero}=\cup^m_{j=1}V^j_{\pinzero}$, $V_{\pinone}=\cup^m_{j=1}V^j_{\pinone}$ and $V_{\eq}=\cup^m_{j=1}\mathcal{V}^j_{\eq}$. 
We wish to argue that $\mathcal{V}$ is admissible for $H$ 
with respect to~$f$. The various disjointness constraints 
in Definition~\ref{def:generalpinfull} are satisfied since 
$H_1,\ldots,H_m$ are disjoint  (using
the fact that each $\mathcal{V}^j$ is admissible for~$H_j$ with respect to~$f_j$).
We have assumed, in the statement of the lemma, that $f$ supports equality.
To show that $\mathcal{V}$ is admissible for $H$ with respect to $f$,
we need to show that if some $f_j$ supports pinning-to-$s$ for some $s\in\{0,1\}$
then so does~$f$. 
This follows from Lemma~\ref{lem:fgpinning} since, by assumption, $f$ simulates $f_j$.
Thus, $\mathcal{V}$ is admissible for $H$ with respect to $f$.

Note that, for every subset $S\subseteq V$ and every assignment $\tau:S\rightarrow \{0,1\}$, it holds that 
\begin{equation}\label{eq:disjointness12}
\mu_{f;H}^{\condV}(\sigma_S=\tau)=\prod^m_{j=1}\mu_{f;H_j}^{\mathrm{cond}(\mathcal{V}_j)}(\sigma_{S\cap V_j}=\tau_{S\cap V_j}).
\end{equation}

Having completed the construction of the desired $H$, to recover \eqref{eq:goaledc}, it remains to specify   $\mathcal{V}^*$   and the vertices $v_1,\hdots,v_{t'}$. 
For each $i\in [t']$, define $U_i$ as in the proof of Lemma~\ref{lem:transitivity}.
Also, let  $U:=\cup_{i\in[t']} U_i$, 
The main idea is that $\mathcal{V^*}$ is the
same as $\mathcal{V}$ except that the sets $U_1,\ldots,U_{t'}$ are added to $V_{\eq}$
because we want to condition on the fact that the variables in each of these sets are equal.
In order to formally specify $\mathcal{V^*} = (V^*_{\pinzero}, V^*_{\pinone}, V^*_{\eq})$
there is a slight technical difficulty because $V^*_{\pinzero}$
and $V^*_{\pinone}$ have to be disjoint  from each other and from all sets in $V^*_{\eq}$.
In order to deal with this (rather unimportant, but technical) detail,
we give an algorithm for defining $\mathcal{V^*}$.
Let $\mathcal{V}^0 = \mathcal{V}$.
Then, for $i=1,\ldots,t'$  define $\mathcal{V}^i=(V^i_{\pinzero},V^i_{\pinone},\mathcal{V}^i_{\eq})$ as follows.
\begin{itemize}
\item Let $\mathcal{V}^i = \mathcal{V}^{i-1}$.
\item Note that no vertex in $U_i$ is in $V_{\pinzero} \cap V_{\pinone}$.
This follows since $I$ is satisfiable (since $g$ is not the always-zero function $\zerof$). 
\item If any vertex in $U_i$ is in $V_{\pinzero}$ then
replace $V^i_{\pinzero}$ with  $V^{i-1}_{\pinzero} \cup U_i$.
\item Otherwise, if any vertex in $U_i$ is in $V_{\pinone}$ then
replace $V^i_{\pinone}$ with $V^{i-1}_{\pinone} \cup U_i$.
\item Otherwise, if $U_i$ does not intersect any sets in $\mathcal{V}^{i-1}_{\eq}$
then replace $\mathcal{V}^i_{\eq}$ with $\mathcal{V}^{i-1}_{\eq} \cup \{U_i\}$.
\item Otherwise, let $W_1,\ldots,W_z$ be the sets in $\mathcal{V}^{i-1}_{\eq}$
that intersect $U_i$ and replace $\mathcal{V}^i_{\eq}$ with
$( \mathcal{V}^{i-1}_{\eq} \setminus \{W_1,\ldots,W_z\} ) \cup \{U_i \cup W_1 \cup \cdots \cup W_z\}$.
\end{itemize}
Finally, let $\mathcal{V^*} = \mathcal{V}^{t'}$.
  Then choose the vertices $v_1,\hdots,v_{t'}$ to be arbitrary vertices in $U_1,\hdots, U_{t'}$, respectively.

It
remains to prove that \eqref{eq:goaledc12} holds. As in the proof of Lemma~\ref{lem:transitivity},
we call an assignment $\tau:U\rightarrow\{0,1\}$ \emph{relevant} if for every $i\in[t']$ there exists $s_i\in\{0,1\}$ such that for every vertex $v\in U_i$, it holds that $\tau(v)=s_i$. For relevant assignments $\tau$, we will refer to the tuple $(s_1,\hdots,s_{t'})$ as the CSP assignment corresponding to $\tau$.  Clearly, for non-relevant $\tau$, we have that  $\mu^{\mathrm{cond}(\mathcal{V}^*)}_{f;H}(\sigma_{U}=\tau)=0$ since $\mathcal{V}^*$ forces equality on each of the sets $U_1,\hdots, U_{t'}$. For all relevant $\tau:U\rightarrow \{0,1\}$, we have that 
\[\mu^{\mathrm{cond}(\mathcal{V}^*)}_{f;H}(\sigma_{U}=\tau)=\frac{\mu_{f;H}^{\condV}(\sigma_{U}=\tau)}{\mu_{f;H}^{\condV}(\sigma^{\eq}_{U_1},\hdots, \sigma^{\eq}_{U_t})}\] and hence
\begin{equation}\label{eq:conditionalexp43312}
\mu^{\mathrm{cond}(\mathcal{V}^*)}_{f;H}(\sigma_{U}=\tau)\propto \mu_{f;H}^{\condV}(\sigma_{U}=\tau) \mbox{ for all relevant $\tau$}.
\end{equation}
Using \eqref{eq:disjointness12}, we have that 
\begin{equation}\label{eq:5r4r54r512}
\mu_{f;H}^{\condV}(\sigma_{U}=\tau)=\prod^m_{j=1}\mu_{f;H_j}^{\mathrm{cond}(\mathcal{V}^j)}(\sigma_{U\cap V_j}=\tau_{U\cap V_j}).
\end{equation}
Note that for every $j\in [m]$ we have $U\cap V_j=\{v_{j,1},\hdots,v_{j,w(j)}\}$ and, hence, \eqref{eq:fsupportsfj12} gives
\begin{equation}\label{eq:epspselect12}
\mu_{f;H_j}^{\mathrm{cond}(\mathcal{V}^j)}(\sigma_{U\cap V_j}=\tau_{U\cap V_j})=\frac{
f_j( \tau(v_{j,1}),\cdots,\tau(v_{j,w(j)}))}
 {|R_{f_j}|}.
\end{equation}
It follows from \eqref{eq:conditionalexp43312}, \eqref{eq:5r4r54r512} and \eqref{eq:epspselect12} that
\begin{equation}\label{eq:epspselect2b12}
\mu^{\mathrm{cond}(\mathcal{V}^*)}_{f;H}(\sigma_{U}=\tau)\propto\prod^m_{j=1}
f_j( \tau(v_{j,1}),\cdots,\tau(v_{j,w(j)}))
  \mbox{ for all relevant $\tau$}.
\end{equation}
For a relevant $\tau:U\rightarrow \{0,1\}$, let $(s_1,\hdots,s_{t'})$ be the CSP assignment corresponding to $\tau$. Then, the product in the r.h.s. of \eqref{eq:epspselect2b12} is 1 iff  $(s_1,\hdots,s_{t'})$ encodes a satisfying assignment of the CSP instance $I$. Since the relevant $\tau:U\rightarrow \{0,1\}$ and assignments to the CSP instance $I$ are in 1-1 correspondence, we obtain \eqref{eq:goaledc12}, as wanted.

This concludes the proof of Lemma~\ref{lem:transitivityb}.
\end{proof}

The following Boolean functions, which were considered in \cite{trichotomy}, will be important in what follows: 
$\delta_0$ and $\delta_1$ (defined in Definition~\ref{def:dopin}), and
 $\XOR$, $\Implies$, $\NAND$, $\Bor$. For convenience, 
 we state the corresponding relations here.
\begin{itemize}
\item $R_{\delta_0}=\{(0)\}$ and $R_{\delta_1}=\{(1)\}$ (these correspond to satisfying assignments of $\neg x$ and $x$, respectively).
\item $R_{\XOR}=\{(0,1),(1,0)\}$ (corresponds to satisfying assignments of $x\neq y$). 
\item $R_{\Implies}=\{(0,0),(0,1), (1,1)\}$ (corresponds to satisfying assignments of $x\Rightarrow y$).
\item $R_{\NAND}=\{(0,0),(0,1), (1,0)\}$ (corresponds to satisfying assignments of $\neg x\vee \neg y$).
\item $R_{\Bor}=\{(0,1), (1,0),(1,1)\}$ (corresponds to satisfying assignments of $ x\vee y$).
\end{itemize}

\subsection{The case of self-dual functions}\label{sec:self-dual}

A Boolean function~$f$ is said to be \emph{self-dual} if, for all $\xb$,
$f(\xb) = f(\overline{\xb})$.
In this section, we 
show (Theorem~\ref{thm:sd-main}, below) that if $f$ is a self-dual Boolean function which is not affine, and $f$~supports perfect equality, then
$f$ simulates a hard function. First,  we establish a useful lemma.

\begin{lemma}\label{lem:SDsimXOR}
Let  $f:\{0,1\}^k\rightarrow \{0,1\}$ be a self-dual function Boolean with $f\neq  \zerof^{(k)}$ and $f(\mathbf{0}) = 0$. Further, suppose that $f$ supports perfect equality. Then, $f$ perfectly simulates $\XOR$.
\end{lemma}
\begin{proof}    
From $f(\mathbf{0}) = 0$ and self-duality, we have that $f(\mathbf{1}) = 0$. Since $f\neq \zerof^{(k)}$, there must be some $\xb\notin\{ \mathbf{0},\mathbf{1}\}$ such that $f(\xb) = 1$. 
By self-duality, we have that $f(\overline{\xb})=1$ as well.
Let $U_0 = \{ i\in [k] \mid x_i=0\}$ and
$U_1 = \{ i \in [k] \mid x_i = 1\}$. 

Since $f$ supports perfect equality, there exists a $k$-tuple hypergraph $H_{\eq}=(V_\eq,\mathcal{F}_\eq)$ and vertices $y,z\in V_\eq$ such that
\begin{equation*} 
\mu_{f;H_{\eq}}(\sigma(y)=\sigma(z)=0)=\mu_{f;H_{\eq}}(\sigma(y)=\sigma(z)=1)=1/2.
\end{equation*} 

Construct the $k$-tuple hypergraph $H$ as follows. First, take a single hyperarc $(v_1,\ldots,v_k)$.
Then, for every $s\in \{0,1\}$ and
every $i,j$  such that $i$ and $j$ are  both in  $U_s$, add a new copy of $H_{\eq}$,
identifying $y$ with $v_i$ and $z$ with $v_j$.
Finally, choose $v_1\in U_0$ and $v_2 \in U_1$.
 Then 
 $$\mu_{f;H}(\sigma(v_1)=0,\sigma(v_2)=1) = \mu_{f;H}(\sigma(v_1)=1,\sigma(v_2)=0) = 1/2,$$
 so $f$ perfectly simulates $\XOR$. 
\end{proof}

\begin{theorem}\label{thm:sd-main}
Suppose that $f$ is a self-dual Boolean function which is not affine and supports perfect equality. Then $f$ simulates a hard function.
\end{theorem}    
 
\begin{proof} 
Let $k$ be the arity of~$f$.
The proof has two cases depending on whether $f(\zeroes)=0$ or $f(\zeroes)=1$. We begin with the case where $f(\mathbf{0})=1$ (we will reduce the proof for the other case to this one).

So, assume first that $f(\zeroes)=1$. Since $f$ is not affine, by applying Item~\ref{it:fixeda} of Lemma~\ref{lem:affine} to $\ab=\zeroes$, we obtain that there exist $\bb,\cb\in \{0,1\}^k$ such that $f(\bb)=f(\cb)=1$ but $f(\bb\oplus \cb)=0$. By self-duality, we also have that $f(\overline{\bb})=f(\overline{\cb})=1$. Note that 
\[\bb\neq\cb,\qquad \bb\neq \zeroes,\ones, \qquad \cb\neq \zeroes,\ones.\]
Indeed, it cannot be the case that $\bb=\cb$ since then $f(\bb\oplus \cb)=f(\mathbf{0})=1$. Analogously, $\bb=\zeroes$ would give that $f(\bb\oplus \cb)=f(\cb)=1$. Similarly, $\bb=\mathbf{1}$ would give that  $f(\bb\oplus \cb)=f(\overline{\cb})=1$. By symmetry between $\bb$ and $\cb$, we have that $\cb\neq \zeroes,\ones$.

Let $w,x,y,z$ be Boolean variables. For $i\in[k]$, let 
\[r_i=\begin{cases} w,&\mbox{ if } b_i=0, c_i=0,\\x ,&\mbox{ if } b_i=0, c_i=1,\\y ,&\mbox{ if } b_i=1, c_i=0,\\z ,&\mbox{ if } b_i=1, c_i=1.\end{cases}\]
Let $V:=\{r_1,\hdots,r_k\}$ (note that $V$ has at most 4 elements). Also, consider the Boolean function $h:\{0,1\}^{|V|}\rightarrow\{0,1\}$ defined by $h=f(r_1,\hdots,r_k)$.

We next study in more detail the function $h$. Observe that
\begin{itemize}
\item $V$ must contain at least one of $x,y$ since $\bb\neq \cb$.
\item $V$ must contain at least one of $w,x$ since $\bb\neq \ones$.
\item $V$ must contain at least one of $w,y$ since $\cb\neq \ones$.
\item $V$ must contain at least one or $y,z$ since $\bb\neq \zeroes$.
\item $V$ must contain at least one of $x,z$ since $\cb\neq \zeroes$.  
\end{itemize}
Thus, the cases to consider are $V=\{w,x,y,z\}$, $|V|=3$, or $V=\{x,y\}$. However, $V=\{x,y\}$ is not possible since then $\bb\oplus\cb=\ones$ and $f(\ones)=1$ (contradicting that $f(\bb\oplus\cb)=0$).  
We now consider the function $h$ (and the corresponding relation $R_h$) in each of the possible cases.

\begin{itemize}
\item Case 1. $V = \{x,y,z\}$. 

Note that $(x,y,z)=(0,0,0)\in R_h$ since $f(\zeroes)=1$. Also, $(0,1,1)\in R_h$ since $f(\bb)=1$. 
Also, $(1,0,1)\in R_h$ since $f(\cb)=1$. By self-duality $(1,1,1)$, $(1,0,0)$, $(0,1,0)$ are also in $R_h$. Then $(x,y,z)=(1,1,0)$ is not in $R_h$ since $f(\bb\oplus \cb)=0$ and by self-duality neither is $(0,0,1)$. So $h(x,y,z)$ is completely determined. Then, for the function $g(x,y) := \sum_z h(x,y,z)$, we have that 
\[g(0,0)=g(1,1)=1 \quad\mbox{ and  }\quad g(0,1)=g(1,0)=2,\]
which is a hard function.
        
\item Case 2. $V = \{w,x,y\}$. This case is similar to Case 1 by switching the spins 0 and 1.
        
\item Case 3. $V = \{w,x,z\}$.       
$(w,x,z)=(0,0,0)$ is in $R_h$ since $f(\zeroes)=1$.
$(0,0,1)$ is in $R_h$ since $f(\bb)=1$.
$(0,1,1)$ is in $R_h$ since $f(\cb)=1$.
By self-duality, $(1,1,1)$, $(1,1,0)$ and $(1,0,0)$ are also in $R_h$.
$(0,1,0)$ is not in $R_h$ since $f(\bb\oplus \cb)=0$.
By self-duality, $(1,0,1)$ is not in $R_h$.
Then, for the function $g(w,z) = \sum_x h(w,x,z)$, we have that  
\[g(0,0)=g(1,1)=1 \quad\mbox{ and  }\quad g(0,1)=g(1,0)=2,\]
which is a hard function.
        
\item Case 4. $V = \{w,y,z\}$. This case follows from Case 3 by switching $\bb$ and $\cb$.
        
\item Case 5. $V = \{w,x,y,z\}$
        
Similarly to the other cases, we have the following tuples in $R_h$:
$(w,x,y,z) = (0,0,0,0)$, $(0,0,1,1)$, $(0,1,0,1)$,  
and their complements
and we know that
$(w,x,y,z)=(0,1,1,0)$ and  its complement are not in $R_h$.
Let $h_0 = h(0,x,y,z)$.  
Let $$C = \{(0,0,1),(0,1,0),(1,0,0),(1,1,1)\}.$$
For every possible subset $S$ of $C$, we have to consider the possibility 
that $R_{h_0} = S \cup \{(0,0,0), (0,1,1),(1,0,1)\}$. This is a lot of cases, but fortunately, some of  them can be combined.
        
\begin{itemize}
\item  Case 5a. $(x,y,z)=(0,1,0)$ is in $S$ but $(1,1,1)$ is not.
Then, for the function $g(x,y) := \sum_w h(w,x,y,x)$, we have
\[g(0,0)=g(1,1)=1 \quad\mbox{ and  }\quad g(0,1)=g(1,0)=2,\]
which is a hard function.
            
\item Case 5b. $(x,y,z) = (1,1,1)$ is in $S$ but $(0,1,0)$ is not.
Then, for the function  $g(w,x) := \sum_y h(w,x,y,x)$, we have
\[g(0,0)=g(1,1)=1 \quad\mbox{ and  }\quad g(0,1)=g(1,0)=2,\]
which is a hard function. 
            
\item Case 5c. $(x,y,z) = (1,0,0)$ is in $S$ but $(1,1,1)$ is not. This case is symmetric to Case 5a.

\item Case 5d. $(x,y,z) = (1,1,1)$ is in $S$ but $(1,0,0)$ is not. This case is symmetric to Case 5b.

\item Case 5e. $(x,y,z)=(0,1,0)$ and $(1,0,0)$ are both in $S$.

Then, for the function  $g(x,y) := \sum_w h(w,x,y,w)$, we have 
\[g(0,0)=g(1,1)=1 \quad\mbox{ and  }\quad g(0,1)=g(1,0)=2,\]
which is a hard function. 
            
\item Case 5f. $S=\emptyset$.
Then, for the function $g(x,y):= \sum_{w,z} h(w,x,y,z)$, we have 
\[g(0,0)=g(1,1)=1 \quad\mbox{ and  }\quad g(0,1)=g(1,0)=2,\]
which is a hard function. 
            
\item Case 5g. $S = \{(0,0,1)\}$.
Then, for the function $g(w,z) := \sum_{x,y} R_h(w,x,y,z)$, we have
\[g(0,0)=g(1,1)=1 \quad\mbox{ and  }\quad g(0,1)=g(1,0)=3,\]
which is a hard function. 
\end{itemize}      
\end{itemize}

It remains to argue that each of the functions $g$ used in Cases 1---5 can be simulated using the function $f$.
This is direct. $\{f\}$ implements the function~$h$ and $\{h\}$ implements the function $g$, so $\{f\}$
implements $g$. Since $f$ supports perfect equality, we can apply Lemma~\ref{lem:transitivity} 
taking $\Gamma = \{f\}$.
Since (trivially) $f$ perfectly simulates $\Gamma$, we find that $f$ perfectly simulates $g$.
This completes the proof for the case where $f(\zeroes)=1$.

We next argue for the case where $f(\zeroes)=0$. Since $f$ is not affine, we have that $f\neq  \zerof^{(k)}$, so there exists $\tb\neq \zeroes$ such that $f(\tb)=1$. Let $S:=\{i\in [k] \mid t_i=1\}$ and note that $S\neq\emptyset$.

Consider the function $f'$ defined by $f'(\xb):=f(\xb\oplus \tb)$ for all $\xb\in \{0,1\}^k$.  Note that 
\begin{itemize} 
\item $f'(\zeroes)=1$, since $f(\tb)=1$.
\item $f'$ is self-dual. Indeed, for $\xb\in \{0,1\}^k$ we have 
\[f'(\overline{\xb})=f(\xb\oplus \ones\oplus \tb)=f(\xb\oplus \tb)=f'(\xb),\]
where the middle equality follows from the self-duality of $f$.
\item $f'$ is not affine.  
Since $f$ is not affine, we know from Lemma~\ref{lem:affine}(1)
that there are $\ab,\bb,\cb$ such that 
$f(\ab)=f(\bb)=f(\cb)=1$ and $f(\ab\oplus \bb\oplus \cb)=0$.
Let $\ab'= \ab\oplus \tb$, $\bb' = \bb \oplus \tb$ and $\cb' = \cb \oplus \tb$.
Then by the definition of~$f'$,
$f'(\ab')=f'(\bb')=f'(\cb')=1$.
But $f'(\ab' \oplus \bb' \oplus \cb') = 
f'(\ab \oplus \bb \oplus \cb \oplus \tb) =
f(\ab \oplus \bb \oplus \cb)=0$, so $f'$ is not affine.

\end{itemize}
By the previous argument, we thus have $\{f'\}$  
implements a hard function $g$.
We  will show that $f$ simulates~$g$. Indeed, observe that the constraint language $\{f,\XOR\}$ implements $f'$ (just apply $\XOR$ to the bits of $f$ which correspond to non-zero entries of the vector $\tb$). Since $f(\zeroes)=0$ and $f$ 
supports perfect equality, by Lemma~\ref{lem:SDsimXOR} we have that $f$ perfectly simulates $\{f,\XOR\}$. 
  Applying
 Lemma~\ref{lem:transitivity}  with $\Gamma = \{f,\XOR\}$ and the $g$ of Lemma~\ref{lem:transitivity} as $f'$, 
 we find that $f$ perfectly simulates $f'$.
Then applying Lemma~\ref{lem:transitivity}  
again with $\Gamma = \{f'\}$, and the $g$ of Lemma~\ref{lem:transitivity} as $g$,
we 
obtain that $f$ simulates the hard function $g$, as wanted.
This concludes the proof of Theorem~\ref{thm:sd-main}.
\end{proof}

\subsection{$\nBIS$-easiness}

The goal of Section~\ref{sec:equalitysec} is to
  prove Theorem~\ref{thm:perfect}. 
 The  required $\nBIS$-easiness results  follows directly from \cite{trichotomy}.
\begin{lemma}[{\cite[Lemma 9]{trichotomy}}]\label{lem:BISeasy}
Let $\Gamma$ be a constraint language such that every relation in $\Gamma$ belongs to $\IM$. Then, $\nCSP{\Gamma}$ is $\nBIS$-easy.\qed
\end{lemma}

\subsection{$\nBIS$-hardness}
We next prove the required $\nBIS$-hardness results (cf. Lemma~\ref{lem:BIShardness} below). We will use the following results from the literature. 
\begin{lemma}[{\cite[Corollary 3]{BoundedBIS}}]\label{lem:BISbounded}
Let $\Delta\geq 6$. It is $\nBIS$-hard to count the number of independent sets in bipartite graphs of maximum degree $\Delta$. \qed
\end{lemma}

The following lemma is from Lemma~13 of \cite{trichotomy}.
We take the lemma from there since we use the notation of~\cite{trichotomy}. However, the
proof is  originally from Lemmas 5.24 and 5.25 of \cite{CKSb}.
 
\begin{lemma}[{\cite{CKSb}}]\label{lem:intermediatepinning}
If $f$ is a  Boolean function that is not self-dual, then $\{f\}$ implements either $\delta_0$ or $\delta_1$.
\end{lemma} 
\begin{proof}
We just need to explain the terminology in \cite[Lemma 13]{trichotomy}. It will be then apparent that Items (i)--(iv) in \cite[Lemma 13]{trichotomy} show that $\{f\}$ implements $\delta_0$ or $\delta_1$. ``$0$-valid" in \cite{trichotomy} means that $\zeroes\in R_f$, ``$1$-valid" means that $\ones\in R_f$ and ``complement-closed" means self-dual.
\end{proof}
    
\begin{lemma}[{\cite[Lemma 15]{trichotomy}, see also \cite{CKSb}}]\label{lem:keytricho}
If $f$ is a Boolean function that is not affine, then $\{f,\delta_0\}$ implements one of $\Bor$, $\Implies$, $\NAND$. The same is true for $\{f,\delta_1\}$. \qed
\end{lemma}
    
We are now ready to show that, for every $f$ which supports perfect equality and is not affine, it holds that, for all sufficiently large $\Delta$, $\HyperSpinfAP$ is $\nBIS$-hard.
\begin{lemma}\label{lem:BIShardness}
Let $f:\{0,1\}^k\rightarrow \{0,1\}$ be a Boolean function which supports perfect equality. Suppose that $f$ is not affine. Then, for all sufficiently large $\Delta$, $\HyperSpinfAP$ is $\nBIS$-hard.
\end{lemma}
\begin{proof}
Assume first that $f$ is self-dual. Then, by Theorem~\ref{thm:sd-main} (note that $f$ is not affine and supports perfect equality by assumption), $f$ perfectly simulates a hard function. By Lemma~\ref{lem:NPinapprox}, we obtain that for all sufficiently large $\Delta$, there exists $c>1$ such that $\HyperSpinf$ is $\NP$-hard. Now, recall that every problem in $\#\mathrm{P}$ admits an FPRAS using an NP-oracle \cite{ValiantVazirani}. Since $\HyperSpinf$ is $\NP$-hard, we can use it as an oracle to obtain an FPRAS for $\nBIS$.

Assume next that $f$ is not self-dual.  By Lemma~\ref{lem:intermediatepinning} we have that $\{f\}$ implements either $\delta_0$ or $\delta_1$. We only need to consider the case where $\{f\}$ implements $\delta_0$, the case of $\delta_1$ follows by just switching the spins 0 and 1. 
First, by Lemma~\ref{lem:transitivity} with $\Gamma=\{f\}$ and $g=\delta_0$, $f$ perfectly simulates $\delta_0$, so
$f$ perfectly simulates $\{f,\delta_0\}$. 
Recall that $f$ is not affine.  By Lemma~\ref{lem:keytricho}, it thus follows that $\{f,\delta_0\}$ implements one of $\Bor$, $\NAND$, $\Implies$.  Using Lemma~\ref{lem:transitivity} again,  it follows that $f$ perfectly simulates one of $\Bor$, $\NAND$, $\Implies$.
        
Note that $\Bor$ and $\NAND$ correspond to hard functions, so when $f$ perfectly simulates either $\Bor$ or $\NAND$, we obtain from Lemma~\ref{lem:NPinapprox} that for all sufficiently large $\Delta$, there exists $c>1$ such that $\HyperSpinf$ is $\NP$-hard. Thus, as in the case of self-dual functions, we may conclude that for all sufficiently large $\Delta$, $\HyperSpinf$ is $\nBIS$-hard.

Thus, it remains to consider the case where $f$ perfectly simulates $\Implies$. By Definition~\ref{def:simulate}, this means that there exists a  $k$-tuple hypergraph $H'=(V',\mathcal{F}')$ and vertices $x,y$ in $H'$ such that
\begin{equation}\label{eq:impliesimplement}
\frac{Z_{00}}{Z_{f;H'}}=\frac{1}{3},\quad \frac{Z_{01}}{Z_{f;H'}}=\frac{1}{3},\quad \frac{Z_{11}}{Z_{f;H'}}=\frac{1}{3},\quad \frac{Z_{10}}{Z_{f;H'}}=0,
\end{equation}
where, for $s_1,s_2\in\{0,1\}$, we denote 
\[Z_{s_1s_2}:=\sum_{\substack{\sigma:V'\rightarrow\{0,1\};\\ \sigma_{x}=s_1,\, \sigma_{y}=s_2}}w_{f;H'}(\sigma).\] 
Let $\Delta'$ be the   degree of $H'$. We will show that for all $\Delta\geq 6\Delta'$, $\HyperSpinf$ is $\nBIS$-hard. 

We will use Lemma~\ref{lem:BISbounded}. In particular, let $G=(V_1\cup V_2, E)$ be a bipartite graph of maximum degree $6$ where $V_1,V_2$ denote the parts of $G$ in its partition. Let $H=(V,\mathcal{F})$ be the  $k$-tuple hypergraph obtained from $G$ as follows. 
Start by putting all of the vertices in $V_1 \cup V_2$ into $V$. Then add additional vertices and hyperarcs as follows.
For every edge $(v_1,v_2)\in E$ such that $v_1\in V_1$ and $v_2\in V_2$, take a distinct copy of $H'$ and identify vertex $x$ in $H'$ with $v_1$ and vertex $y$ in $H'$ with $v_2$. Note that $V_1\cup V_2\subseteq V$ and that the   degree of $H$ is $6\Delta'$.

Let $\mathcal{I}_G$ denote the set of independent sets of $G$. Then, we claim that 
\begin{equation}\label{eq:qwe23ree3}
Z_{f;H}=|\mathcal{I}_G|\cdot (Z_{f;H'}/3)^{|E|}.
\end{equation}
Before proving \eqref{eq:qwe23ree3}, note  that an oracle call to $\HyperSpinfAP$ for $\Delta\geq 6\Delta'$ with input $H$ and relative error $\epsilon>0$ yields via \eqref{eq:qwe23ree3} an estimate for the number of independent sets in  bipartite graphs of maximum degree 6 which is within relative error $\epsilon$ from the true value. Thus, using Lemma~\ref{lem:BISbounded}, we obtain an AP-reduction from $\nBIS$ to  $\HyperSpinfAP$ for all $\Delta\geq 6\Delta'$, as wanted. 
 
To show \eqref{eq:qwe23ree3}, let $\sigma:V\rightarrow \{0,1\}$ be an assignment such that $w_{f;H}(\sigma)>0$. The copies of $H'$ ensure that for every edge $(v_1,v_2)\in E$ such that $v_1\in V_1$ and $v_2\in V_2$ it holds that either $\sigma(v_1)\neq 1$ or $\sigma(v_2)\neq 0$. Thus, the set $(\sigma^{-1}(1)\cap V_1)\cup (\sigma^{-1}(0)\cap V_2)$ is an independent set of $G$. Conversely, for every independent set $I$ of $G$, consider 
\[\Omega_I=\{\sigma:V\rightarrow \{0,1\}\mid \sigma_{I\cap V_1}=\ones,\ \sigma_{I\cap V_2}=\zeroes,\  \sigma_{V_1\backslash I}=\zeroes,\ \sigma_{V_2\backslash I}=\ones\}.\]
Then, using \eqref{eq:impliesimplement}, we have that the number of assignments $\sigma\in \Omega_I$ such that $w_{f;H}(\sigma)>0$ is equal to
\[\prod_{(v_1,v_2)\in E}Z_{\sigma(v_1)\sigma(v_2)}=(Z_{f;H'}/3)^{|E|}.\]
Summing this over all $I\in \mathcal{I}_G$, we obtain \eqref{eq:qwe23ree3}, thus completing the proof of Lemma~\ref{lem:BIShardness}.
\end{proof}

\subsection{NP-hardness}

In this section, we show (Lemma~\ref{lem:NPhardness} below)
that if $f$ supports perfect equality, and it is not affine, and is not in $\IM$ 
then,  for all sufficiently large $\Delta$, there exists $c>1$ such that $\HyperSpinf$ is $\NP$-hard.
 To do this, we need  
 some preparation. The  ideas behind the following lemma are essentially from~\cite{trichotomy}.
  
\begin{lemma} \label{lem:delta0delta1}
If $f$ perfectly simulates $\Implies$, then $f$ simulates $\{\delta_0,\delta_1\}$ (not necessarily perfectly).
\end{lemma}
\begin{proof}
We first show that $f$ supports both pinning-to-0 and pinning-to-1. Since $f$ perfectly simulates $\Implies$, there exists a $k$-tuple hypergraph $H$ and vertices $v_1,v_2$ in $H$ such that 
\begin{equation}\label{eq:qwsxazdredcf6789}
\mu_{00}=1/3,\quad \mu_{01}=1/3,\quad \mu_{11}=1/3, \quad \mu_{10}=0,
\end{equation}
where, for $s_1,s_2\in\{0,1\}$, we denote $\mu_{s_1s_2}:=\mu_{f;H}(\sigma_{v_1}=s_1,\sigma_{v_2}=s_2)$.
        
Note that 
\begin{equation*}
\begin{gathered}
\mu_{f;H}(\sigma_{v_1}=0)=\mu_{00}+\mu_{01}, \quad \mu_{f;H}(\sigma_{v_1}=1)=\mu_{10}+\mu_{11},\\
\mu_{f;H}(\sigma_{v_2}=0)=\mu_{00}+\mu_{10}, \quad \mu_{f;H}(\sigma_{v_2}=1)=\mu_{01}+\mu_{11}.
\end{gathered}
\end{equation*}
It follows that 
\begin{equation}\label{eq:tgbg565565}
\begin{gathered}
\mu_{f;H}(\sigma_{v_1}=0)=2/3,\quad  \mu_{f;H}(\sigma_{v_1}=1)=1/3,\\
\mu_{f;H}(\sigma_{v_2}=0)=1/3, \quad \mu_{f;H}(\sigma_{v_2}=1)=2/3.
\end{gathered}
\end{equation}
Then, using \eqref{eq:tgbg565565}, we obtain that  $H$ and its vertices $v_1,v_2$ satisfy the assumptions of Lemma~\ref{lem:gadgets} ($v_1$ satisfies Item~\ref{it:pinning0} and $v_2$ Item~\ref{it:pinning1}) and hence we obtain that $f$ supports pinning-to-0 and pinning-to-1. 

To conclude that $f$ simulates $\{\delta_0,\delta_1\}$, consider the $k$-tuple hypergraph $H$ as above  and consider the conditional distribution $\mu_{f;H}^{\mathcal{V}_0}$ where we pin the vertex $v_2$ to 0 (this is allowed since $f$ supports pinning-to-$0$). Then,
\[\mu_{f;H}^{\mathcal{V}_0}(\sigma(v_1)=0)=1, \quad \mu_{f;H}^{\mathcal{V}_0}(\sigma(v_1)=1)=0\]
so $f$ simulates $\delta_0$. Analogously, by pinning the vertex $v_1$ to 1, we also obtain that $f$ simulates $\delta_1$, concluding the proof.
\end{proof}
\begin{lemma}[{\cite[Proof of Lemma 19]{trichotomy}}]\label{lem:secondpiece}
If $f$ is a  Boolean function that is not in $IM_2$, then $\{f,\Implies,\delta_0,\delta_1\}$ implements either $\Bor$ or $\NAND$. \qed
\end{lemma}

\begin{lemma}\label{lem:NPhardness}
Let $f:\{0,1\}^k\rightarrow \{0,1\}$ be a function which supports perfect equality. Suppose that $f$ is not affine and is not in $\IM$. Then, for all sufficiently large $\Delta$, there exists $c>1$ such that $\HyperSpinf$ is $\NP$-hard.
\end{lemma}
\begin{proof} 
The proof is similar in structure to the proof of Lemma~\ref{lem:BIShardness}.
 
Assume first that $f$ is self-dual. Then, by Theorem~\ref{thm:sd-main} (note that $f$ is not affine and supports perfect equality by assumption), $f$ perfectly simulates a hard function. By Lemma~\ref{lem:NPinapprox}, we obtain that for all sufficiently large $\Delta$, there exists $c>1$ such that $\HyperSpinf$ is $\NP$-hard. 

Assume next that $f$ is not self-dual.   By Lemma~\ref{lem:intermediatepinning} we have that $\{f\}$ implements either $\delta_0$ or $\delta_1$. We only need to consider the case where $\{f\}$ implements $\delta_0$, the case of $\delta_1$ follows by switching the spins 0 and 1. 
First, by Lemma~\ref{lem:transitivity} with $\Gamma=\{f\}$ and $g=\delta_0$, $f$ perfectly simulates $\delta_0$, so
$f$ perfectly simulates $\{f,\delta_0\}$. 
Recall that $f$ is not affine.  By Lemma~\ref{lem:keytricho}, it thus follows that $\{f,\delta_0\}$ implements one of $\Bor$, $\NAND$, $\Implies$.  Using Lemma~\ref{lem:transitivity} again,  it follows that $f$ perfectly simulates one of $\Bor$, $\NAND$, $\Implies$.
$\Bor$ and $\NAND$ correspond to hard functions, so when $f$ perfectly simulates either $\Bor$ or $\NAND$, we obtain from Lemma~\ref{lem:NPinapprox} that for all sufficiently large $\Delta$, there exists $c>1$ such that $\HyperSpinf$ is $\NP$-hard. Thus, it remains to consider the case where $f$ perfectly simulates $\Implies$.   

Since $f$ perfectly simulates $\Implies$, by Lemma~\ref{lem:delta0delta1}, we obtain that  $f$ simulates $\{\delta_0,\delta_1\}$. Thus, $f$ simulates $\{f,\Implies,\delta_0,\delta_1\}$. By Lemma~\ref{lem:secondpiece}, using that $f$ is not in $IM_2$, we have that $\{f,\Implies,\delta_0,\delta_1\}$ implements either $\Bor$ or $\NAND$. By Lemma~\ref{lem:transitivityb},  
we thus obtain that $f$ simulates either $\Bor$ or $\NAND$. Hence, as above, we can use Lemma~\ref{lem:NPinapprox} to conclude that for all sufficiently large $\Delta$, there exists $c>1$ such that $\HyperSpinf$ is $\NP$-hard. 

This concludes the proof. 
\end{proof}    
    
\subsection{Proof of Theorem~\ref{thm:perfect}}
We are ready to prove Theorem~\ref{thm:perfect}, which we restate here for convenience.
\begin{thmperfect}
\statethmperfect{}
\end{thmperfect}
\begin{proof} 
Item 1 is a consequence of Lemma~\ref{lem:BISeasy} and Lemma~\ref{lem:BIShardness}. Item 2 is a consequence of Lemma~\ref{lem:NPhardness}.
\end{proof}

\section{Proof of Theorem~\ref{thm:main}}\label{sec:proofofmain}
In this section, we combine the pieces to prove Theorem~\ref{thm:main}. 

We will need the following lemma.
\begin{lemma}\label{lem:conc}
Let $f_1:\{0,1\}^{k_1}\rightarrow \{0,1\}$ and $f_2:\{0,1\}^{k_2}\rightarrow \{0,1\}$ be Boolean functions such that $f_1$ is not affine and $f_2$ is not in $\IM$. Then, the function $f$ defined by
$f(\xb,\yb)=f_1(\xb)f_2(\yb)$ is neither affine nor does it belong to $\IM$.
\end{lemma}
\begin{proof} 
We first prove that $f$ is not affine. Since $f_1$ is not affine, by Item~\ref{it:abc} of Lemma~\ref{lem:affine}, there exist $\xb^{(1)},\xb^{(2)},\xb^{(3)}\in R_{f_1}$ such that  $\xb^{(1)}\oplus \xb^{(2)}\oplus \xb^{(3)}\notin R_{f_1}$. Let $\yb$ be such that $f_2(\yb)=1$ (such a $\yb$ exists, otherwise $f_2$ would belong to $\IM$). 

For each $i=1,2,3$, consider the vector $\zb^{(i)}$ of  length $k_1+k_2$ obtained by concatenating the vectors $\xb^{(i)}$ and $\yb$. Since $\xb^{(i)}\in R_{f_1}$ and $\yb\in R_{f_2}$, we have that $f(\zb^{(i)})=f_1(\xb^{(i)})f_2(\yb)=1$, so $\zb^{(i)}\in R_{f}$ for $i=1,2,3$. Observe that $f(\zb^{(1)}\oplus \zb^{(2)}\oplus \zb^{(3)})=f_1(\xb^{(1)}\oplus\xb^{(2)}\oplus\xb^{(3)})f_2(\yb)=0$, so $\zb^{(1)}\oplus \zb^{(2)}\oplus \zb^{(3)}\notin R_f$. Thus,
\[\zb^{(1)},\zb^{(2)},\zb^{(3)}\in R_f \mbox{ but } \zb^{(1)}\oplus\zb^{(2)}\oplus\zb^{(3)}\notin R_f,\]
so by Item~\ref{it:abc} of Lemma~\ref{lem:affine}, we have that $f$ is not affine.

We next show that $f$ does not belong to $\IM$. Since $f_2\notin \IM$, by Lemma~\ref{lem:IM2}, there exist $\yb^{(1)},\yb^{(2)}\in R_{f_2}$ such that either $\yb^{(1)}\vee \yb^{(2)}\notin R_{f_2}$ or $\yb^{(1)}\wedge \yb^{(2)}\notin R_{f_2}$. Assume that $\yb^{(1)}\vee \yb^{(2)}\notin R_{f_2}$, the other case is completely analogous and actually follows by duality (switching the spins 0 and 1).  Let $\xb$ be such that $f_1(\xb)=1$ (such an $\xb$ exists, otherwise $f_1$ would be affine).

For each $i=1,2$, consider the vector $\wb^{(i)}$ of  length $k_1+k_2$ obtained by concatenating the vectors $\xb$ and $\yb^{(i)}$. Since $\xb\in R_{f_1}$ and $\yb^{(i)}\in R_{f_2}$, we have that $f(\wb^{(i)})=f_1(\xb)f_2(\yb^{(i)})=1$, so $\zb^{(i)}\in R_{f}$ for $i=1,2$. Observe that $f(\wb^{(1)}\vee \wb^{(2)})=f_1(\xb)f_2(\yb^{(1)}\vee\yb^{(2)})=0$, so $\wb^{(1)}\vee \wb^{(2)}\notin R_f$. Thus,
\[\wb^{(1)},\wb^{(2)}\in R_{f} \mbox{ but } \wb^{(1)}\vee \wb^{(2)}\notin R_f,\]
so by Lemma~\ref{lem:IM2}, we have that $f$ does not belong to $\IM$.

This concludes the proof.
\end{proof}
\begin{thmmain}
\statethmmain{}
\end{thmmain}
\begin{proof}
We consider each of the three cases. 
\begin{enumerate}
\item If every function in $\Gamma$ is affine (cf. Definition~\ref{def:affine}), 
then $Z_{I}$ can be computed exactly in polynomial time using Gaussian elimination. 
This was already noted in the exact-counting dichotomy of Creignou and Hermann~\cite{CH}.

\item  Suppose that $\Gamma\subseteq \IM$ and that $\Gamma$ includes a function $f$ which is not affine. By 
the unbounded-degree $\nBIS$-easiness result of~\cite{trichotomy}, 
which is stated here as
Lemma~\ref{lem:BISeasy}, it follows that
for all positive integers $\Delta$,
$\nCSPd{\Gamma}$ is $\nBIS$-easy. Clearly, every instance of $\nNRCSPd{\Gamma}$ is an instance of $\nCSPd{\Gamma}$, from which we obtain that $\nNRCSPd{\Gamma}$ is $\nBIS$-easy as well.

If $f$ supports perfect equality, then by Theorem~\ref{thm:perfect}, 
for all sufficiently large~$\Delta$, the problem
$\HyperSpinfAP$ is $\nBIS$-hard. As we noted in Section~\ref{sec:prelim}, 
the problem $\HyperSpinfAP$ is equivalent to $\nNRCSPd{\{f\}}$ from which we obtain that $\nNRCSPd{\Gamma}$ is $\nBIS$-hard as well. Note that $\nNRCSPd{\Gamma}$ is a restricted version 
of $\nCSPd{\Gamma}$ (the restriction being that constraints may not repeat variables),
so it follows immediately that $\nCSPd{\Gamma}$ is $\nBIS$-hard.

There is a final case 
that does not arise if $\nBIS$ is not $\NP$-hard to approximate, but we include it to make the proof complete.
In particular, if $f$ does not support perfect equality, then   
by Theorem~\ref{thm:asym-main}, it simulates a hard function. So, by 
Lemma~\ref{lem:NPinapprox},  for all sufficiently large $\Delta$, there exists $c>1$ such that 
$\HyperSpinf$ is $\NP$-hard. As observed in the proof of Lemma~\ref{lem:BIShardness}, this implies  that $\HyperSpinfAP$ is $\nBIS$-hard. Then, as in the previous case, we obtain that $\nNRCSPd{\Gamma}$ and $\nCSPd{\Gamma}$ are $\nBIS$-hard.

\item Suppose that there  are functions $f_1,f_2\in \Gamma$ such that $f_1$ is not affine and $f_2$ is not in $\IM$ (it might be the case that $f_1=f_2$). Then, consider the function $f(\xb,\yb)$ defined by $f(\xb,\yb)=f_1(\xb)f_2(\yb)$. By Lemma~\ref{lem:conc}, we have that $f$ is neither affine nor does it belong to $\IM$. 

Thus, if $f$ supports perfect equality,  then by Theorem~\ref{thm:perfect}, 
 for all sufficiently large $\Delta$, there exists $c>1$ such that 
 $\HyperSpinf$ is $\NP$-hard, which  is equivalent to saying that $\nNRCSPdc{\{f\}}$ is $\NP$-hard.
 Now note that  there is an easy reduction from 
 $\nNRCSPdc{\{f\}}$ to $\nNRCSPdc{\Gamma}$ ---
 given an instance $I$ of $\nCSPdc{\{f\}}$, every constraint involving~$f$
 is re-written as two constraints involving~$f_1$ and~$f_2$. 
 Thus,  $\nNRCSPdc{\Gamma}$ is also $\NP$-hard. Since $\nNRCSPd{\Gamma}$ is a restricted version 
of $\nCSPd{\Gamma}$, we have that $\nCSPdc{\Gamma}$ is $\NP$-hard as well.
 
 Otherwise, 
by Theorem~\ref{thm:asym-main}, $f$ simulates a hard function. So, by 
Lemma~\ref{lem:NPinapprox},  
for all sufficiently large $\Delta$, there exists $c>1$ such that 
$\HyperSpinf$ is $\NP$-hard.  As in the previous  paragraph, this implies that 
$\nNRCSPdc{\Gamma}$ and $\nCSPdc{\Gamma}$ are   $\NP$-hard. \qedhere
\end{enumerate}
\end{proof}

\bibliographystyle{plain}
\bibliography{\jobname}
    
\end{document}